\documentclass[aps,prd,reprint,showpacs,floatfix,footinbib,superscriptaddress,longbibliography]{revtex4-1}
\usepackage{bbding}
\usepackage{amsmath}
\usepackage{amssymb}
\usepackage{latexsym}
\usepackage{graphics}
\usepackage{graphicx}
\usepackage{slashed}
\usepackage{color}
\usepackage{comment}
\usepackage{cancel}
\usepackage[normalem]{ulem}

\usepackage[disable,colorinlistoftodos,backgroundcolor=white,bordercolor=steelblue,linecolor=steelblue]{todonotes}

\usepackage{bm}
\usepackage[colorlinks=true,citecolor=cyan,urlcolor=blue,bookmarks=true,bookmarksopen=true,bookmarksnumbered=true,bookmarksopenlevel=3]{hyperref}

\definecolor{airforceblue}{rgb}{0.36, 0.54, 0.66}
\definecolor{steelblue}{rgb}{0.27, 0.51, 0.71}
\definecolor{amber}{rgb}{1.0, 0.49, 0.0}

\makeatletter
\newsavebox\myboxA
\newsavebox\myboxB
\newlength\mylenA

\newcommand*\xoverline[2][0.75]{%
    \sbox{\myboxA}{$\m@th#2$}%
    \setbox\myboxB\null
    \ht\myboxB=\ht\myboxA%
    \dp\myboxB=\dp\myboxA%
    \wd\myboxB=#1\wd\myboxA
    \sbox\myboxB{$\m@th\overline{\copy\myboxB}$}
    \setlength\mylenA{\the\wd\myboxA}
    \addtolength\mylenA{-\the\wd\myboxB}%
    \ifdim\wd\myboxB<\wd\myboxA%
       \rlap{\hskip 0.5\mylenA\usebox\myboxB}{\usebox\myboxA}%
    \else
        \hskip -0.5\mylenA\rlap{\usebox\myboxA}{\hskip 0.5\mylenA\usebox\myboxB}%
    \fi}
\makeatother

\usepackage{booktabs,amsmath}
\usepackage{amssymb}
\usepackage{bm}
\usepackage{amsmath}
\usepackage{array}
\usepackage{mathtools}
\usepackage{amsthm}
\usepackage{newlfont}
\usepackage[utf8]{inputenc}
\usepackage{esdiff}
\usepackage{multirow}
\usepackage{graphics}
\usepackage{graphicx}
\usepackage{ae,aecompl,color}

\usepackage[bookmarks=true,bookmarksopen=true,bookmarksnumbered=true,bookmarksopenlevel=3]{hyperref}

\makeatletter
\DeclareRobustCommand*{\bfseries}{%
  \not@math@alphabet\bfseries\mathbf
  \fontseries\bfdefault\selectfont
  \boldmath
}
\makeatother

\begin{document}
\newcommand{\bd}{\begin{document}}
\newcommand{\ed}{\end{document}}
\newcommand{\bc}{\begin{center}}
\newcommand{\ec}{\end{center}}
\newcommand{\bfr}{\begin{flushright}}
\newcommand{\efr}{\end{flushright}}
\newcommand{\lt}{\left}
\newcommand{\rt}{\right}
\newcommand{\vs}{\vspace}
\newcommand{\hs}{\hspace}
\newcommand{\beq}{\begin{equation}}
\newcommand{\eeq}{\end{equation}}
\newcommand{\lb}{\linebreak}
\newcommand{\pb}{\pagebreak}
\newcommand{\mb}{\makebox}
\newcommand{\fb}{\framebox}
\newcommand{\mc}{\multicolumn}
\newcommand{\ben}{\begin{enumerate}}
\newcommand{\een}{\end{enumerate}}
\newcommand{\bit}{\begin{itemize}}
\newcommand{\eit}{\end{itemize}}
\newcommand{\un}{\underline}
\newcommand{\lefq}{\lefteqn}
\newcommand{\ba}{\begin{array}}
\newcommand{\ea}{\end{array}}
\newcommand{\beqa}{\begin{eqnarray}}
\newcommand{\eeqa}{\end{eqnarray}}
\newcommand{\beqas}{\begin{eqnarray*}}
\newcommand{\eeqas}{\end{eqnarray*}}
\newcommand{\bfg}{\begin{figure}}
\newcommand{\efg}{\end{figure}}
\newcommand{\bds}{\begin{displaymath}}
\newcommand{\eds}{\end{displaymath}}
\newcommand{\btb}{\begin{tabbing}}
\newcommand{\etb}{\end{tabbing}}
\newcommand{\para}{\parallel}
\newcommand{\pad}{\partial}
\newcommand{\nn}{\nonumber}
\newcommand{\la}{\leftarrow}
\newcommand{\ra}{\rightarrow}
\newcommand{\lgla}{\longleftarrow}
\newcommand{\lgra}{\longrightarrow}
\newcommand{\La}{\Leftarrow}\newcommand{\Ra}{\Rightarrow}
\newcommand{\Lra}{\Leftrightarrow}
\newcommand{\Lgla}{\Longleftarrow}
\newcommand{\Lgra}{\Longrightarrow}
\newcommand{\lan}{\langle}
\newcommand{\ran}{\rangle}
\renewcommand{\a}{\alpha}
\renewcommand{\b}{\beta}
\newcommand{\g}{\gamma}
\newcommand{\G}{\Gamma}
\renewcommand{\d}{\delta}
\newcommand{\eps}{\epsilon}
\newcommand{\Th}{\Theta}
\newcommand{\s}{\sigma}
\newcommand{\lam}{\lambda}
\newcommand{\D}{\Delta}
\newcommand{\vare}{\varepsilon}
\newcommand{\pr}{\prime}
\newcommand{\ro}{\rho}
\newcommand{\nab}{\nabla}
\newcommand{\m}{\mu}
\newcommand{\n}{\nu}
\newcommand{\Sg}{\Sigma}
\newcommand{\p}{\pi}
\newcommand{\R}{I\!\!R}
\newcommand{\om}{\omega}
\newcommand{\Om}{\Omega}
\newcommand{\ze}{\zeta}
\newcommand{\vart}{\vartheta}
\newcommand{\tri}{\triangle}
\newcommand{\f}{\frac}
\newcommand{\iny}{\infty}
\newcommand{\pro}{\propto}
\renewcommand{\arraystretch}{1.25}
\title{Angular momentum quantum backflow in the noncommutative plane. } 
%
%
\author{\textsc{Valentin Daniel Paccoia}}
\affiliation{Dipartimento di Fisica e Geologia, Università degli Studi di Perugia, Via A. Pascoli, 06123, Perugia, Italy}
\email[Email:]{valentindaniel.paccoia@studenti.unipg.it}
\author{\textsc{Orlando Panella}}
\affiliation{Istituto Nazionale di Fisica Nucleare, Sezione di Perugia, Via A.~Pascoli, I-06123 Perugia, Italy}
\email[({\bf Corresponding Author}) Email: ]{orlando.panella@pg.infn.it }
\author{\textsc{Pinaki Roy}}
\affiliation{Atomic~Molecular~and~Optical~Physics~Research~Group, Advanced Institute of Materials Science, Ton Duc Thang University, Ho~Chi~Minh~City, Vietnam}
\email[Email:]{pinaki.roy@tdtu.edu.vn}
\affiliation{Faculty of Applied Sciences, Ton Duc Thang University, Ho~Chi~Minh~City,~Vietnam}

\date{\today}

\begin{abstract}
We study the quantum backflow problem in the noncommutative plane. In particular, we have considered a charged particle with and without an oscillator interaction with noncommuting momentum operators and examined angular momentum backflow in each case and how they differ from each other. We also propose a probability associated with the occurence of angular momentum backflow and investigate whether or not the probability depends on a physical parameter, namely the magnetic field.
\end{abstract}


\maketitle

\section{Introduction}
Quantum backflow is a striking but still relatively not well known quantum effect for which, 
\todo[noline,size=\tiny]{Point n.8}
given a state containing only postive momentum components, the probability of observing the particle to the right of a given reference point ($x=0$ for instance), may actually decrease over time. This amounts to saying  that there is a flow of the probability density in the  direction opposite to that of the momentum. In other words, it means that a right-moving particle can actually move to the left. The effect was first discovered by Allcock in his works on arrival time and it was noted that the probability current could be negative for states consisting only of positive momenta \cite{allcock1,allcock2,allcock3}.

A detailed investigation of the problem was carried out and an upper bound to the amount of probability that can flow in a direction opposite to momentum was found \cite{bm94}. This limit $c_{bm}$ has a numerically computed value of about $0.04$ and the most surprising fact is that it is a dimensionless value, independent of any physical parameter. Because of that it has been considered as “a new quantum number”. This may cause some problems in the naive classical limit  $\hbar\to 0$ as it has been observed how there are different systems for which the maximum amount of backflow becomes dependent on some physical parameters. 
\todo[noline,size=\tiny]{Point n.9 }
However it has been shown \cite{yearsley2012}, in a one dimensional setting, that with a more realistic approach introducing  quasi-projectors $\theta_\sigma(\hat{x})$~\footnote{The smoothed quasi-projector $\theta_\sigma(\hat{x}) = \int_{0}^{\infty} dy \delta_\sigma(\hat{x}-y)$ is defined in terms a smoothed (over a length scale $\sigma$) Dirac-$\delta$: $\delta_\sigma(\hat{x}-y) =\frac{1}{\sqrt{2\pi\sigma^2}} \exp\left[-\frac{(\hat{x}-y)^2}{\sigma^2}\right]$.} of the position operator smoothed, over a length scale $\sigma$, instead of the ideal projector  $\theta(\hat{x})$, the  limit $\hbar\to 0$  reproduces correctly the classical behavior, i.e. no backflow. On the other hand, in the case of a Dirac particle backflow was found to depend on some physical parameters~\cite{melloy2,melloy3} but not in a way that explains the classical limit. 
\todo[noline,size=\tiny]{Point n.5 }
See also~\cite{Su:2018aa,Ashfaque:2019aa} for a discussion of the quantum backflow in the Dirac equation of spin 1/2 free particles.

There have been attempts to improve the value $c_{bm}$ \cite{eveson,penz2005}. In particular, in ref \cite{penz2005} an operator associated with the backflow problem was found and was used to improve upon the value of $c_{bm}$. 
However, \todo[noline,size=\tiny]{Point n.10 Refs. here?} although the problem of finding the exact eigenstate corresponding to the upper bound of $c_{bm}$ (maximum backflow) has still not been solved analytically, there are plenty of constructions of backflowing states. 

\todo[noline,size=\tiny]{Point n.11 } In an interesting approach \cite{berry10}, the backflow problem was studied using  non normalizable wave functions and it was found that in relation to the superoscillations a constraint regarding spatial extension exists. By measuring the fraction of the $x$-axis subject to backflow, the probability of finding the particle in one one these regions, and their temporal evolution was found. Interestingly quantum backflow has also been studied in different contexts like decay of metastable states \cite{vanDijk2019aa}, in multiparticle systems~\cite{Barbier2020:aa}, appearance of classically forbidden probability flux \cite{Goussev2019:aa} etc.
\todo[noline,size=\tiny]{Point n.6 } In another recent work~\cite{goussev:2020aa}  the author discusses the relationship between  quantum backflow and quantum reentry (QR), the effect in which a wave packet evolving from a localized spatial region partially returns to this region in the absence of external forces, providing a unifying treatment of the two effects.



It may be noted that in interacting systems, the backflow problem can be studied in different ways. \todo[noline,size=\tiny]{Point n.7 } For example, quantum backflow has been studied in the context of scattering states in ref.~\cite{Bostelmann:2017aa} where  it has been shown that those features of the probability operator in the quantum backflow in the case of no interactions are also found when interactions are present. The most important properties of the quantum backflow are stable against the introduction of interaction potentials even strong ones. 

In another example~\cite{strange,Goussev2020:ab}, considering an electron in a magnetic field, the backflow problem has been \todo[noline,size=\tiny]{Point n.12 } formulated in terms of an effective angular momentum defined for states given as a  superposition of eigenstates and written out the in phase-amplitude form.
It was shown \cite{strange} that in certain regions of space the effective angular momentum can be directed in a direction opposite to that of the wave function's components. \todo[noline,size=\tiny]{Point n.13 }  These results are very interesting since they show that while the usual momentum quantum backflow is related to the uncertainty relation between the position coordinate and the momentum  \cite{strange}, the angular momentum backflow relates to the uncertainty relations between the polar (azimuthal) coordinate and the $L_z^{\text{can}}$ component of the angular momentum operator. This could mean that similar effects could be found for other models
\todo[noline,size=\tiny]{Point n.14 } where additional interactions are present.

During the past decade or so studies on quantum gravity and string theory indicate that space may be noncommutative in nature \cite{Douglas:2001aa,Connes:1998aa,Seiberg:1999aa}. In order to test the effect of space noncommutativity several quantum mechanical models e.g., harmonic oscillator \cite{Bellucci:2001aa,Smailagic:2002ab}, central field problems \cite{Gamboa:2002aa,Gamboa:2001ab}, hydrogen problem \cite{PhysRevLett.86.2716,Chaichian2004} etc., have been studied within the framework of noncommutative quantum mechanics. In all these cases attempts were made to determine the effect of noncommutativity by finding the dependence of some observable like energy on the parameter(s) of noncommutativity. In some other cases the effect of noncommutativity on phenomena like chirality phase transition \cite{PhysRevA.90.042111}, Hall effect in Dirac matter like graphene \cite{bertolami,Duval_2001,doi:10.1063/1.1504484} etc., have been studied.  However, in view of the fact that no conclusive evidence regarding noncommutative nature of space or momenta has yet been conclusively established, it is of interest to find new models where this effect may eventually be detected. In this context it may be noted that in recent years experiments have been proposed to detect quantum backflow in Bose-Einstein condensate \cite{PhysRevA.87.053618} as well as in the field of optics \cite{eliezer2018observation}. 

In the present paper our objective is to study quantum backflow problem on the the noncommutative plane. To be more specific, we shall consider two models: (1) The first one is a noncommutative analogue of a charged particle in the presence of a homogeneous magnetic field, (2) A noncommutative oscillator in a homogeneous magnetic field. In both the models we shall study angular momentum backflow \cite{strange} and examine to what extent noncommuting nature of the momentum operators affects quantum backflow. In this context it may be noted that the second model is a more general one from which the first one can be obtained by setting the oscillator frequency equal to zero. Nevertheless we have treated them separately as the first one has a commutative analogue \cite{strange} with which we may compare our results and the second one is a completely new one. Secondly, it also helps us to find out the difference in backflow pattern when an additional interaction is present. Finally, we shall also make an attempt to quantify angular momentum backflow by defining a suitable probability associated with it. The organization of the paper is as follows: in Section \ref{formu} we formulate the model(s) on the noncommutative plane and obtain the solutions; in Section \ref{sssec:str} we study angular momentum backflow in a noncommutative setting; in Section \ref{bfprob} we define a probability associated with angular momentum backflow and discuss some of its features; finally Section \ref{con} is devoted to a conclusion.

\section{Noncommutative charged particle subject to an oscillator in a magnetic field}\label{formu}
To begin with we note that the Hamiltonian $H_{\text{NC}}$ for a particle of charge $q$  in the non-commutative plane in the presence of a homogeneous magnetic field subject to an oscillator potential \todo[noline,size=\tiny]{Point n.18 } has the same functional form as the one in the commutative plane. Thus the Hamiltonian  $H_{\text{NC}}$ is taken to be of the form:
\beq\label{h1}
H_{\text{NC}}=\frac{1}{2\mu}\left(\bm{\hat p}-\frac{q}{c}\bm{{\hat A}}\right)^2+\frac{1}{2} \mu \omega^2 \left(\hat{x}^2+\hat{y}^2\right)\,,
\eeq
where $c$ is the velocity of light, and $\mu$ is the particle's mass. We choose the vector potential to be analogous to the one in the commutative plane producing a constant magnetic field along the $z$ axis as $\bm{B}=B\,\bm{\hat{k}}$, $\bm{\hat{k}}$ being the $z$-axis unit vector:
\beq
\bm{\hat{A}}=(-B{\hat y}/2,B{\hat x}/2)\, .
\eeq
The commutation relation between the non-commuting coordinates and momenta are given by\todo[noline,size=\tiny]{Point n.15 }\textcolor{steelblue}{ \cite{Bertolami:2005aa}}
\beq
\begin{array}{ll}
\label{NCEQ}
[{\hat x},{\hat y}]=i\theta, &\quad[{\hat p}_x,{\hat p}_y]=i\eta,\cr [{\hat x}_i,{\hat p}_j]=i\hbar(1+\f{\theta\eta}{4\hbar^2})\delta_{ij},&\quad\theta,\eta \in \mathbb{R}\,.
\end{array}
\eeq
Then for an electron, charge $q=-e$,  the above non commutative Hamiltonian reads
\beq
H_{\text{NC}}= \frac{1}{2\mu} \left({\hat p_x}-\frac{eB}{2c}{\hat y}, {\hat p_y}+\frac{eB}{2c}{\hat x}\right)^2+\frac{1}{2} \mu \omega^2 \left(\hat{x}^2+\hat{y}^2\right).
\label{h2}
\eeq
Using the commutation relations in Eq.~(\ref{NCEQ}) we \todo[noline,size=\tiny]{Point n.16 } obtain:
\begin{eqnarray}
\left({\hat p_x}-\frac{eB}{2c}{\hat y} \right)^2 &=& {\hat p_x}^2 +\left(\frac{eB}{2c}\right)^2 {\hat y}^2-\frac{eB}{c}{\hat y}{\hat p_x},\\
\left( {\hat p_y}+\frac{eB}{2c}{\hat x}\right)^2 &=&{\hat p_y}^2 +\left(\frac{eB}{2c}\right)^2 {\hat x}^2+\frac{eB}{c}{\hat x}{\hat p_y}.
\end{eqnarray}
It is now necessary to express the non-commuting coordinates and momenta in terms of commuting ones. This can be achieved using the Seiberg-Witten map~\todo[noline,size=\tiny]{Point n.17 }\cite{Kokado:2004aa} and the transformations are given by:
\begin{equation}
\ba{lcl}
{\hat x}=\displaystyle {x-\f{\theta}{2\hbar}p_y,~~~~{\hat p}_x=p_x+\f{\eta}{2\hbar}y}\,,\\
{\hat y}=\displaystyle{y+\f{\theta}{2\hbar}p_x,~~~~{\hat p}_y=p_y-\f{\eta}{2\hbar}x}\label{rel}\,,
\ea
\end{equation}
where $(x,y)$ and $(p_x,p_y)$ denote commuting coordinates and momenta. Now using the relations in Eq.~\eqref{rel} the Hamiltonian in Eq.~\eqref{h2} can be written as:
\begin{eqnarray}
H=&&\displaystyle\frac{1}{2\mu}\left[\left(1-\frac{eB\theta}{4c\hbar}\right)p_x+\left(\frac{\eta}{2\hbar}-\frac{eB}{2c}\right)y\right]^2\nonumber\\
&&\displaystyle+\frac{1}{2\mu}\left[\left(1-\frac{eB\theta}{4c\hbar}\right)p_y-\left(\frac{\eta}{2\hbar}-\frac{eB}{2c}\right)x\right]^2\nonumber\\
&&\displaystyle+\frac{1}{2}\mu\omega^2\left[\left(x-\frac{\theta}{2\hbar}p_y\right)^2+\left(y+\frac{\theta}{2\hbar}p_x\right)^2\right]
\end{eqnarray}
The above expression can be more conveniently written in terms of the following frequencies: 
\begin{equation}
\tilde{\omega}=\frac{eB}{2\mu c},\, \qquad \omega_\theta = \frac{2\,\hbar}{\mu\theta},\,\qquad \omega_\eta=\frac{\eta}{2\hbar \mu}.
\end{equation}
Then we find:
\begin{equation}
 H_{\text{NC}}= \alpha \frac{p_x^2+p_y^2}{2\mu}  +\frac{1}{2} \mu \beta \left(x^2 +y^2\right) +\gamma \left(xp_y-yp_x\right),
 \label{HNC1}
\end{equation}
where constants $\alpha,\beta,\gamma$ are given by:
\begin{subequations}
\begin{align}
    \alpha=&\left(1-\frac{\tilde\omega}{\omega_\theta}\right)^2 +\frac{\omega^2}{\omega_\theta^2},\\
    \beta=&\left(\tilde\omega-\omega_\eta\right)^2+\omega^2,\label{abcb}\\
    \gamma=&\left(\tilde\omega-\omega_\eta\right)\left(1-\frac{\tilde\omega}{\omega_\theta}\right)-\frac{\omega^2}{\omega_\theta}.\label{abcc}
\end{align}
\label{abc}
\end{subequations}
Upon recognizing that the last term in Eq.~(\ref{HNC1}) involves involves the $z$ component of the angular momentum operator $L_z^{\text{can}}=(\bm{r}\times\bm{p})_z = xp_y-yp_x$, the non-commutative Hamiltonian can be written in the following form: 
\begin{equation}
 H_{\text{NC}}= \sqrt{\alpha}\left\{ \frac{p_x^2+p_y^2}{2\frac{\mu}{\sqrt{\alpha}}}  +\frac{1}{2} \frac{\mu}{\sqrt{\alpha}} \beta \left(x^2 +y^2\right) +\frac{\gamma}{\sqrt{\alpha}} L_z^{\text{can}}\right\},
 \label{HNC2}
\end{equation}
and therefore defining
\begin{subequations}\label{mf}
\begin{align}
    M=\frac{\mu}{\sqrt{\alpha}},\\
    \Omega= \sqrt{\beta},
\end{align}
\end{subequations}
we can finally write:
\begin{equation}
  H_{\text{NC}}=  \sqrt{\alpha}\left[
   {H}^\text{2D}_\circ +\frac{\gamma}{\sqrt{\alpha}} L_z^{\text{can}}\right] \,,
   \label{HNC3}
\end{equation}
where:
\begin{equation}
 {H}^\text{2D}_\circ =   \frac{p_x^2+p_y^2}{2M} +\frac{1}{2}M\Omega^2 (x^2+y^2).
\end{equation}
 We see therefore that the Hamiltonian in Eq.~\eqref{HNC1} can be related to ${H}^\text{2D}_\circ$, the Hamiltonian of a well known \emph{and exactly solvable} non-relativistic system --that of a two dimensional isotropic (or circular) harmonic  oscillator (of frequency $\Omega$ and mass $M$). The eigenfunctions and eigenvalues of this system are well known~\cite{Flugge:1974aa} and can be readily used to solve the non commutative Hamiltonian of Eq.~\eqref{HNC3} since the angular momentum operator $L_z^{\text{can}}$ commutes  with ${H}^{\text{2D}}_\circ$. Thus, the complete set of eigenfunctions and the corresponding eigenvalues for the NC Hamiltonian are identified  by a radial quantum number $n =0, 1, 2, \cdots$ and the angular momentum quantum number $m=0,\pm1, \pm2, \cdots$~\cite{Flugge:1974aa} and are given by:
\begin{subequations}\label{unrM}
\begin{align}
\label{unrMa}
\psi_{n,m}(r,\varphi)\, &= \, C_{n,m} \, r^{|m|} \, e^{-\frac{r^2}{4a_B^2}}\, \times\nonumber\\ \phantom{x}&\phantom{xxxx}\! \!\!\phantom{F}_1F_1 (-n, |m|+1;\frac{r^2}{2a_B^2}) \, e^{im\varphi}\,,\\
\label{unrMb}
{\varepsilon}_{n,m} \,&=\,  \hbar\Omega \left( |m| + 1 + 2n \right)\sqrt{\alpha}+\hbar m \gamma\,,
\end{align}
\end{subequations}
where $a_B=\displaystyle\sqrt{\frac{\hbar}{2M\Omega}}$ and $C_{n,m}$ are normalization constants that can be easily computed as:
\begin{equation}
\label{Cnorm}
C_{n,m}\, = \,
\frac{a_B^{-(|m|+1)}}{\sqrt{\pi2^{|m|+1}}}  \, \frac{\sqrt{\Gamma(|m| +1 +n)}}{\Gamma(|m|+1)\sqrt{\Gamma(n+1)}}\, .
\end{equation}
By setting $\omega=0$ we obtain the results for the non-commutative charged particle in a magnetic field, which we will discuss as the commutative counterpart has already been studied \cite{strange}. At this point we digress a little to point out some features of the spectrum. We note that the spectrum is non degenerate because of the presence of the last term on the r.h.s. of Eq.~(\ref{unrMb}).  However one may easily verify that the usual degeneracy pattern is recovered when $\omega=0,~\theta=0$. We shall see later that non degeneracy of the spectrum has interesting consequences.
\\

So far we have described the systems keeping both space as well as momentum noncommutativity. \todo[noline,size=\tiny]{Point n.19 } It may be noted that momentum noncommutativity ($\eta\neq 0$) produces a magnetic field like effect in the commutative plane while space non-commutativity ($\theta\neq 0$), although it affects other parameters like the mass, does not produce such a magnetic field like effect on the commutative plane. Thus we shall henceforth consider only noncommuting momentum operators. Note that in this case the Hamiltonian can be written in the form:
\begin{equation}
H= \frac{1}{2\mu}(\bm{p}-\bm{A})^2+\frac{1}{2}\mu\omega^2(x^2+y^2),
 \end{equation}
 where $\bm{A}=\frac{1}{2}((-B+\frac{c\eta}{e\hbar})y,(B-\frac{c\eta}{e\hbar})x)$.
In the case without the oscillator part, a critical value is found for the magnetic field $ B_{\text{cr}}$ for which the Hamiltonian becomes that of a free particle of mass $M$ given by (\ref{mf}). Imposing the condition $\Omega=0$ one immediately gets the value of the critical field:
 \begin{equation}
 \label{bcrit}
     B_{\text{cr}}=\frac{\eta c}{\hbar e}.
 \end{equation}
On the other end it is easily seen that when the oscillator is present ($\omega \neq 0$) then the equation $\Omega=0$ does not  have (real) solutions implying that in this case there is no critical value of the magnetic field.\\

\section{Angular momentum backflow on the noncommutative plane}\label{sssec:str}
 Before studying angular momentum backflow, let us note that the current density for a state described by the wave-function $\Psi$ is given by:
 \begin{eqnarray}
  \bm{j}&=&\frac{\hbar}{2Mi}(\Psi^*\nabla \Psi-\Psi\nabla \Psi^*)-\frac{e}{Mc}{\mathbf A}\Psi^*\Psi\nonumber\\
 &=&\bm{j}_1-\bm{j}_2.
\label{current}
 \end{eqnarray}
It may be noted that in terms of cylindrical polar coordinates, the vector potential can be written as
\begin{equation}
A_r=0,~~A_\varphi=\frac{1}{2}\left(B-\frac{c\eta}{e\hbar}\right)r.    \end{equation}
\todo[noline,size=\tiny]{Point n.20 }Next, we discuss a very important concept, namely that of effective angular momentum. The effective angular momentum $\ell_{\text{eff}}= \hbar\, m_{\text{eff}}$ is defined through \cite{strange}:
\begin{subequations}\label{meff}
\begin{align}
     m_{\text{eff}}(r,\varphi) &= \frac{\partial}{\partial\varphi} \text{Arg}~\Psi(r,\varphi)=\text{Im} \frac{\partial\Psi(r,\varphi)/\partial\varphi}{\Psi(r,\varphi)}\label{meffa}\\&=\left(\frac{\mu\,  r}{\hbar}\right)\,\frac{\bm{e_\varphi}\cdot\bm{j}_1(r,\varphi)}{\rho},\label{meffb} 
 \end{align}
\end{subequations}
where $\Psi(r,\varphi)$ denotes the wavefunction in the amplitude-phase form:
\begin{equation}
    \Psi(r,\varphi)=\sqrt{\rho(r,\varphi)} \, \exp\left[{\displaystyle{i \int_0^\varphi \, d\varphi'\, m_{\text{eff}}(r,\varphi')}}\right],
    \end{equation}
$\bm{e_\varphi} \cdot \bm{j}_1(r,\varphi)$ being the probability current along the azimuthal direction and $\rho=|\Psi(r,\varphi)|^2$ is the probability density \cite{Flugge:1974aa}. 

\subsection{Model without Oscillator Interaction: \texorpdfstring{$\omega=0$}{omega=0}}

Now we consider a specific example of a (normalized) wavefunction which is a simple sum of three eigenfunctions, as given by Eq.~\eqref{unrM}. We consider only the $n = 0$ states with three different, non positive, magnetic quantum numbers (-2,-1,0). For the sake of comparison this is analogous to what has been done in \cite{strange} and allows to compute explicitly the local effective magnetic quantum number $m_{\text{eff}}(r,\varphi)$ in a simple closed form. Using Eq.~(\ref{unrM})  we obtain the state $\Psi_0(r,\varphi)$ as:
\begin{eqnarray}\label{wave1}
    \Psi_0(r,\varphi)&=&  \frac{1}{\sqrt{3}}  \sum_{m=0}^2 c_m(r)\, e^{-im\varphi}\nonumber \\&=& \frac{1}{\sqrt{3}} \bigl[c_0(r)+c_1(r)e^{-i\varphi}+c_2(r)e^{-2i\varphi}\bigr],
\end{eqnarray}
where the coefficients $c_m(r)$ are given by
\begin{equation}
\label{coefficients}
  c_m(r)=C_{0,m}r^{|m|}e^{-\frac{r^2}{4a_B^2}},
\end{equation}
and we note that $C_{0,m}=C_{0,-m}=C_{0,|m|}$ as given by Eq.~\eqref{Cnorm}. Note that the state in Eq.~\eqref{wave1} is a simple sum of three eignestates from Eq.~\eqref{unrMa} with $n=0$ and angular momentum $L_z^{\text{can}}= -\hbar (0,1,2)$ respectively.

It is important to note that, in the model without oscillator interaction an infinite degeneracy for fixed $n$ and non positive $m$ is present, therefore, for our wavefunction, time dependence occurs only in the form of an immaterial global phase. Time dependence for the model with oscillator interaction will be discussed in the next sub-section (\ref{withoscillator}).

From Eq.~\eqref{meffa} we can determine the effective value of the quantum number $m_{\text{eff}}(r,\varphi)$ for the superposition in Eq.~\eqref{wave1}:
\begin{widetext}
\begin{equation}\label{meff1}
    m_{\text{eff}}(r,\varphi)=-\frac{c_1^2+2c_2^2+c_0c_1\cos{\varphi}+3c_1c_2\cos{\varphi}+2c_0c_2\cos{2\varphi}}{c_0^2+c_1^2+c_2^2+2c_0c_1\cos{\varphi}+2c_1c_2\cos{\varphi}+2c_0c_2\cos{2\varphi}}.
\end{equation}
\end{widetext}
In order to have $m_{\text{eff}}>0$, we need to look for values of $(r,\varphi)$ where $m_{\text{eff}}$ passes through zero. This leads to
\begin{equation}\label{meffeq}
    \sqrt{2}r\cos^2{\!\!\varphi}+\left(\frac{a_B}{\sqrt{2}}+\frac{3r^2}{4a_B}\right)\cos{\varphi}+\frac{r^3}{4a_B^2}+\left(\frac{1-\sqrt{2}}{2}\right)r=0.
\end{equation}
In the non-commutative setting the form of Eq.~(\ref{meffeq}) remains the same for both  models and the differences lie in the different dependence of frequency and mass contained in $a_B$ on the parameter $\eta$.
\begin{figure}[b!]
    \includegraphics[width=0.9\linewidth]{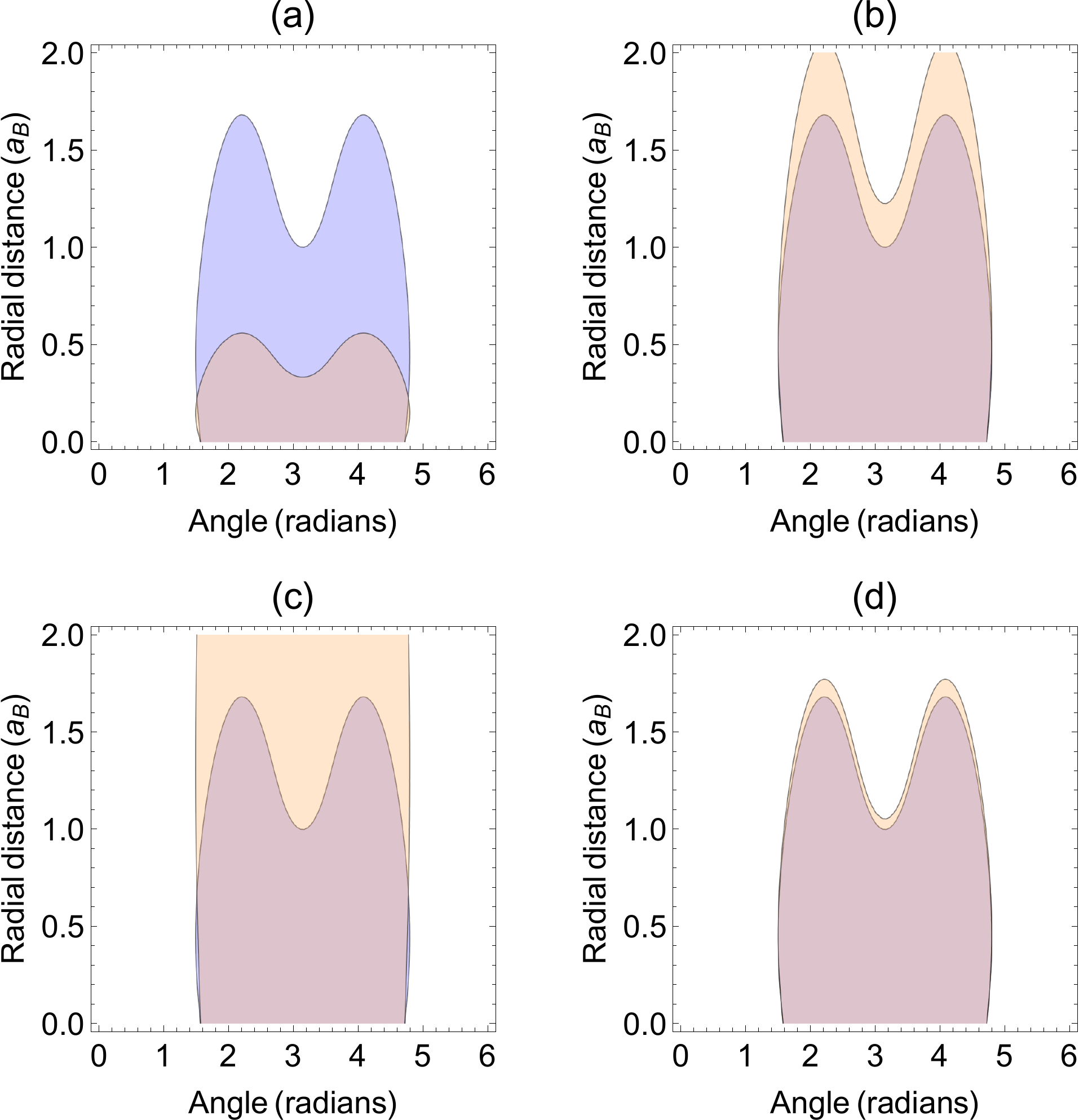}
    \caption{Plot of the region (shaded area) where $m_{\text{eff}}>0$ in the ($r,\varphi $) plane for the commutative case (blue) and the non-commutative case with $\eta/m_e^2c^2=10^{-25}$ (orange) with $\omega=0$. The values of the magnetic field are $B/B_{\text{cr}}=0.1$ (a), $B/B_{\text{cr}}=0.6$ (b), $B/B_{\text{cr}}=0.9$ (c), $B/B_{\text{cr}}=10$ (d). 
    {The radial distance, $r$, is in units of $a_B$.}}
    \label{fig:FIG1}
\end{figure}
To make an easy com\-pa\-ri\-son with different models, it is convenient t87ùùo describe the system in dimensionless units by using the parameter $a_B$ as a unit of length, thus, in general, we will use the value $a_B$ as in the commutative setting. It is important to note that, as the magnetic field changes, $a_B$ also changes, which means that the unit of length will change as well. We recall that in some previous works a bound on the parameter $\eta$ was obtained by comparing non-commutative predictions with measurements and is given by \cite{bertolami,Bertolami:2005aa}:
\begin{equation}
    \sqrt{\eta}\lesssim 2.26\  \frac{\mu\text{eV}}{c}\, .
    \label{etabound}
\end{equation}
 \begin{figure}[t]
    \centering
    \includegraphics[width=0.9\linewidth]{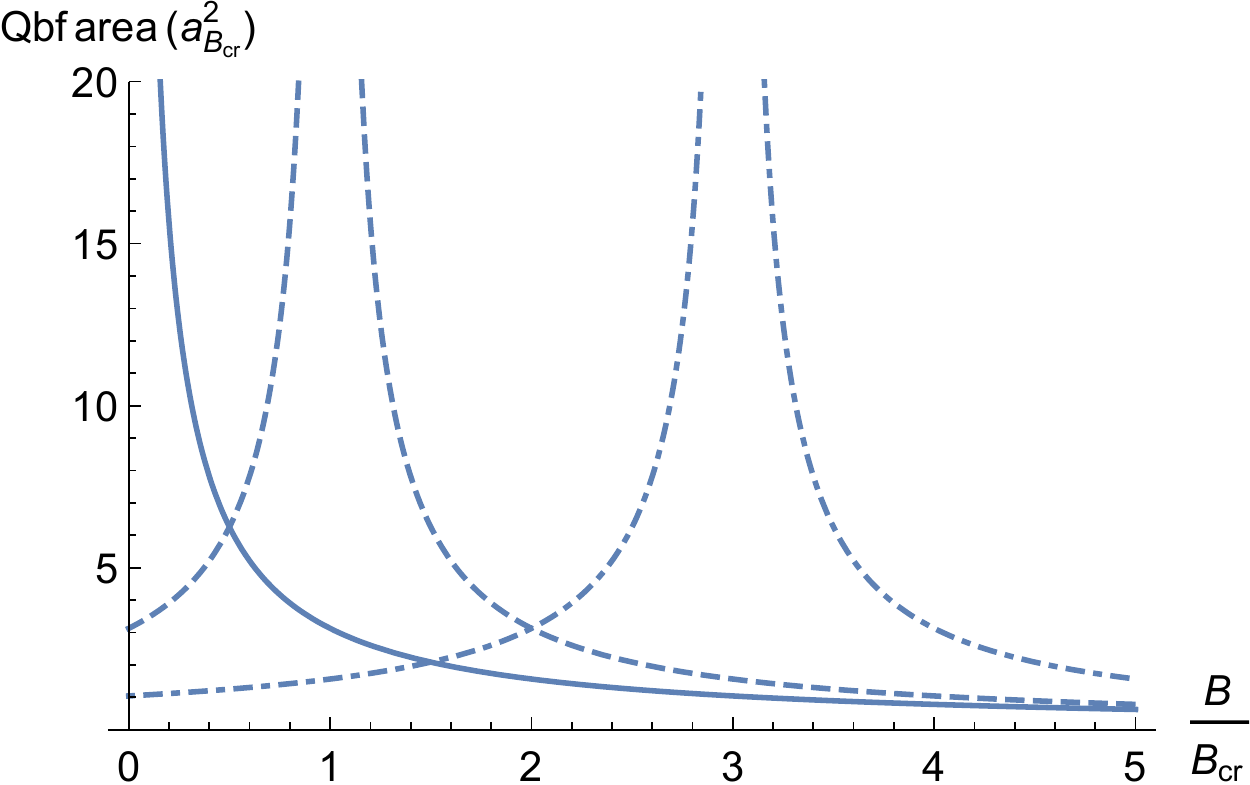}
    \caption{Plot of the backflow area 
    {(in units of $a_{B_{cr}}^2$)} as a function of the magnetic field  for a state with three components as in Eq.~\protect\eqref{wave1} and for two values of the parameter $\eta$ with $\omega=0$, $\eta_1=10^{-25}m_e^2c^2$ (dashed line), $\eta_2=3\times10^{-25}m_e^2c^2$ (dot-dashed line) compared to the commutative result (solid line).}
    \label{fig:FIG2}
\end{figure}
\begin{figure}[t]
     \includegraphics[width=0.9\linewidth]{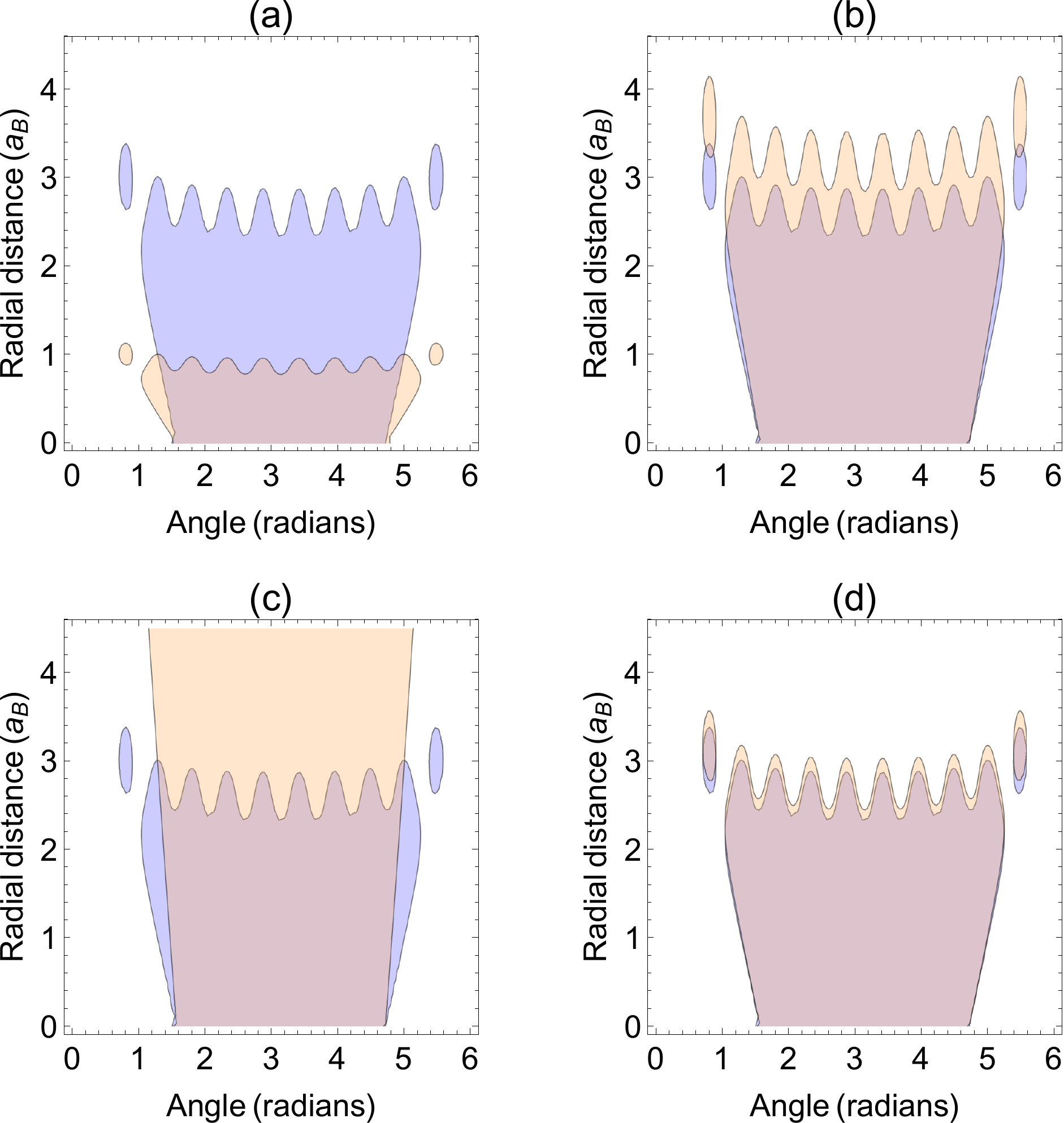}
    \caption{Plot of the region (shaded area) where $m_{\text{eff}}>0$ in the ($r,\varphi $) plane for the commutative case (blue) and the non-commutative case with $\eta/m_e^2c^2=10^{-25}$ (orange) with $\omega=0$ for a state with $N=11$ components as in Eq.~\protect\eqref{psin}. The values of the magnetic field are:   (a) $B/B_{\text{cr}}=0.1$,  (b) $B/B_{\text{cr}}=0.6$,  (c) $B/B_{\text{cr}}=0.9$,  (d) $B/B_{\text{cr}}=10$. 
    {The radial distance, $r$,  is in units of $a_B$.}}
    \label{fig:FIG3}
\end{figure}

Thus we set $\eta/m_e^2c^2=10^{-25},~m_e$ being the electron mass and it is consistent with (\ref{etabound}).
Next, choosing different values of the magnetic field around the critical value,  $B_{\text{cr}}$, given in Eq.~(\ref{bcrit}), we determine the regions of quantum backflow and the results are given in  Fig.(\ref{fig:FIG1}).

As expected from Eq. (\ref{meffeq}), the backflow area changes only radially. To understand how the parameters contribute to the value of $a_B$ we have to study Eq. (\ref{mf}) with $\omega=0$. Approaching the value $B_{cr}$, backflow extends radially to infinity, since $\Omega$ tends to zero and thus $a_B$ tends to infinity, in the limit of small magnetic field the dominant contribution to $\Omega$ is due to the parameter $\eta$ which is constant, in the limit of large magnetic field we return to the commutative case. This is evident if we look at Fig.(\ref{fig:FIG2}) where the backflow area has been plotted as a function of the magnetic field. The unit of area has been chosen as a fixed value of $a_{B_{cr}}^2$ where $\eta=10^{-25}\,m_e^2c^2$ to be able to make a comparison between areas at different values of the magnetic field. The magnetic field is expressed in units $B_{cr}$ where $\eta=10^{-25}m_e^2c^2$.

Next we consider a (normalized) superposition of eigenfunctions consisting of a larger number, $N$, --instead of just 3 as in Eq.~\eqref{wave1}-- of angular momentum eigenstates 
\todo[noline,size=\tiny]{Point n.21 } of the form
 \begin{equation}
   \Psi_{0}(r,\varphi)=\frac{1}{\sqrt{N}}\sum_{m=0}^{N-1} c_m(r)e^{-im\varphi}
     \label{psin}
 \end{equation}
that is a linear combination of $N$ components with non positive angular momentum  ($L_z^{\text{can}}=-\hbar m \le 0) $. The coefficients $c_m(r)$ are defined as in Eq.~\eqref{coefficients}. In the $\omega=0$ 
setting we get the results shown in Fig.~(\ref{fig:FIG3}) by repeating the same steps we already showed for the case of three eigenfunctions. We note that the behaviour of this system is exactly the same as in the previous case, and thus we come to the conclusion that, for $n=0$, the equation that gives us the points $(r,\varphi)$ for which $m_{\text{eff}}$ passes through zero, can always be written in the form $f(r/a_B,\varphi)=0$, where $f$ will be given by a function similar to that given in Eq.~\eqref{meffeq} but with a larger number of terms since we are here considering a state $\Psi$ with $N=11$ components, as in Eq.~\eqref{psin}. Therefore compared to the results in \cite{strange} --solid line in Fig.~(\ref{fig:FIG3})--, the shape of the region with backflow is similar to the noncommutative case (dashed line) but with the radial extension being mostly affected when changing the value of the magnetic field. 


\subsection{\label{withoscillator}Model with Oscillator Interaction: \texorpdfstring{$\omega\neq 0$}{Omega neq 0}}

\begin{figure}[b!]
    \includegraphics[width=0.9\linewidth]{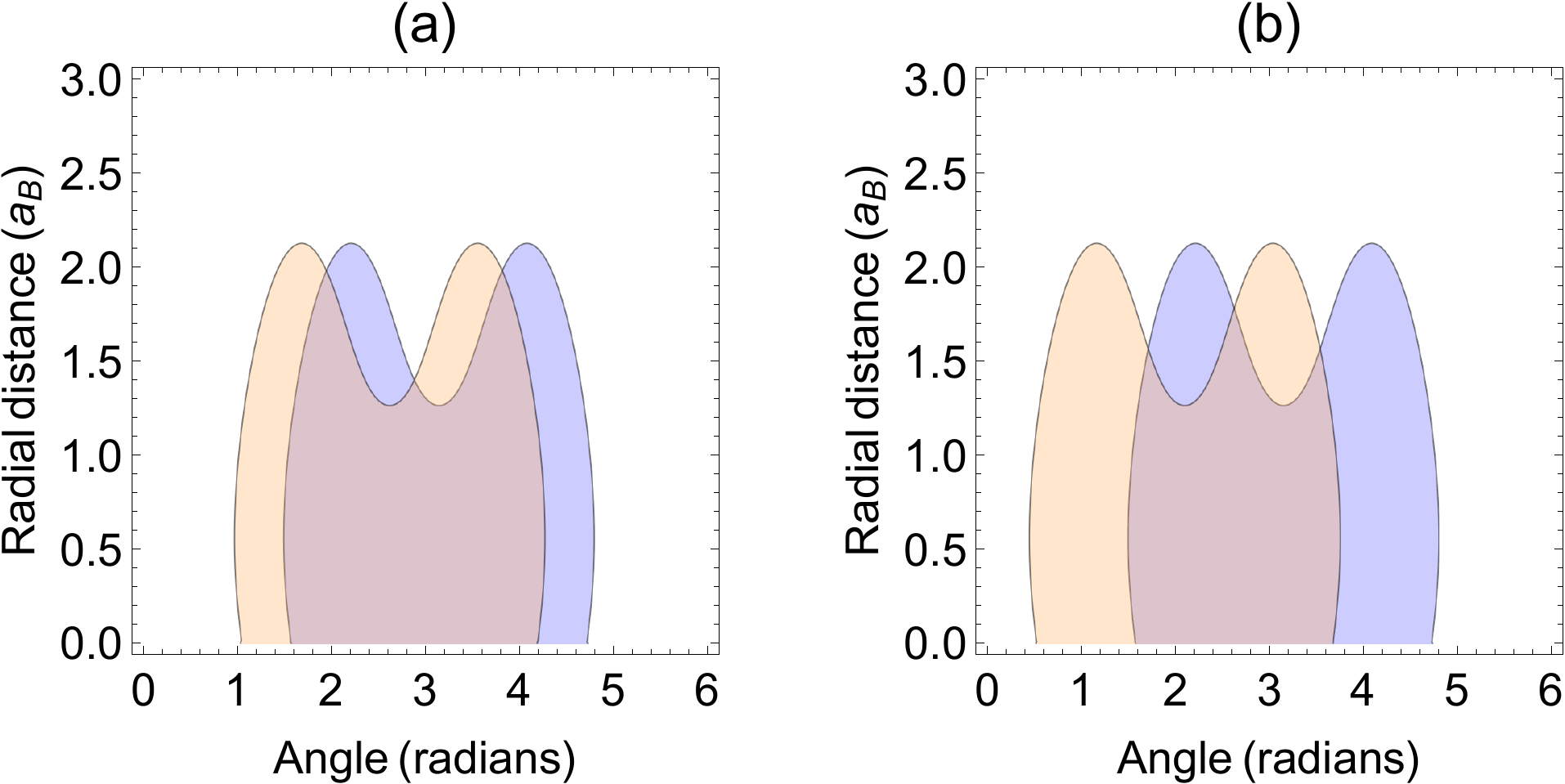}
    \caption{Plot of the region (shaded area) where $m_{\text{eff}}>0$ for two values of time (orange), compared to the result at $t=0$ (blue), in the ($r,\varphi $) plane. The values of time are $t_a=\frac{t_{2\pi}}{12}$ and $t_b=\frac{t_{2\pi}}{6}$, where $t_{2\pi}=2\pi\frac{\hbar}{\epsilon_{-1}-\epsilon_0}$ is the time relative to a $2\pi$ shift, which depends on the magnetic field. Here we consider the case with the oscillator ($\omega=0.8\, \omega_\eta$ with $\eta/m_e^2c^2=10^{-25}$) for a single value of the magnetic field. 
    {The radial distance, $r$, is in units of $a_B$.}}
    \label{fig:FIG4}
\end{figure}

 In this section we shall analyze the effect of the oscillator interaction on the angular momentum backflow. It may be noted that when an oscillator interaction is present, a critical field can not be defined as before: At least in the sense that there is no value of  $B$ that makes the Hamiltonian $H$ in Eq.~\eqref{HNC2} that of a free particle. The value $B_{\text{cr}}$ in this case simply minimizes the quantity $\beta$ in Eq.~(\ref{abcc}) and thus $\Omega$. We can therefore say that the field $B_{\text{cr}}$ minimizes in this instance the interaction. 
 
 Nonetheless for the sake of convenience we shall make our plots with respect to $B/B_{cr}$.
 
 First of all we note that the infinite  degeneracy for fixed $n$ and non positive values of $m$ is no longer present when considering the oscillator interaction. Indeed from Eq.~\eqref{unrMb} we see that, when $\theta = 0$, $\omega_\theta\to \infty$, $\alpha \to 1$, $\Omega=\sqrt{\beta}\to \gamma$ and so when $m\ge 0$ there is degenracy with respect to $m$. Clearly such degeneracy is lifted when the oscillator frequency $\omega\ne 0$ because in this case $\Omega=\sqrt{\beta} \ne \gamma$. This in turn    introduces a time dependence in the wavefunction of Eq.~\eqref{wave1} because each component will have now a different energy eigenvalue. This in turn leads to a time dependence of the effective magnetic quantum number $m_{\text{eff}}$  and leads to:
\begin{widetext}
\begin{equation}
    \small
    m_{\text{eff}}(r,\varphi,t)=-\frac{c_1^2+2c_2^2+c_0c_1\cos{\left[\varphi+\left(\epsilon_{-1}-\epsilon_0\right)\frac{t}{\hbar}\right]}+3c_1c_2\cos{\left[\varphi+\left(\epsilon_{-2}-\epsilon_{-1}\right)\frac{t}{\hbar}\right]}+2c_0c_2\cos{\left[2\varphi+\left(\epsilon_{-2}-\epsilon_0\right)\frac{t}{\hbar}\right]}}{c_0^2+c_1^2+c_2^2+2c_0c_1\cos{\left[\varphi+\left(\epsilon_{-1}-\epsilon_0\right)\frac{t}{\hbar}\right]}+2c_1c_2\cos{\left[\varphi+\left(\epsilon_{-2}-\epsilon_{-1}\right)\frac{t}{\hbar}\right]}+2c_0c_2\cos{\left[2\varphi+\left(\epsilon_{-2}-\epsilon_0\right)\frac{t}{\hbar}\right]}}.
    \label{mefft}
\end{equation}
\end{widetext}
where $\epsilon_m=\epsilon_{0,m}$.  The time dependence reduces to a translation of the angle in $m_{\text{eff}}$.
 This effect is shown explicitly in Fig.(\ref{fig:FIG4}) where we plot the backflow region in the ($r,\varphi$) plane for  two different time values -- see caption of figure --. We show the region for  $t=0$ (delimited by the dashed curve) which corresponds exactly to the time independent case of Eq.~\eqref{meff1} while the solid curve delimits the $\varphi$ shifted one correpsonding to Eq.~\eqref{mefft}. The area of the quantum backflow area will be the same.
 
\begin{figure}[b]
    \centering
   \includegraphics[width=0.9\linewidth]{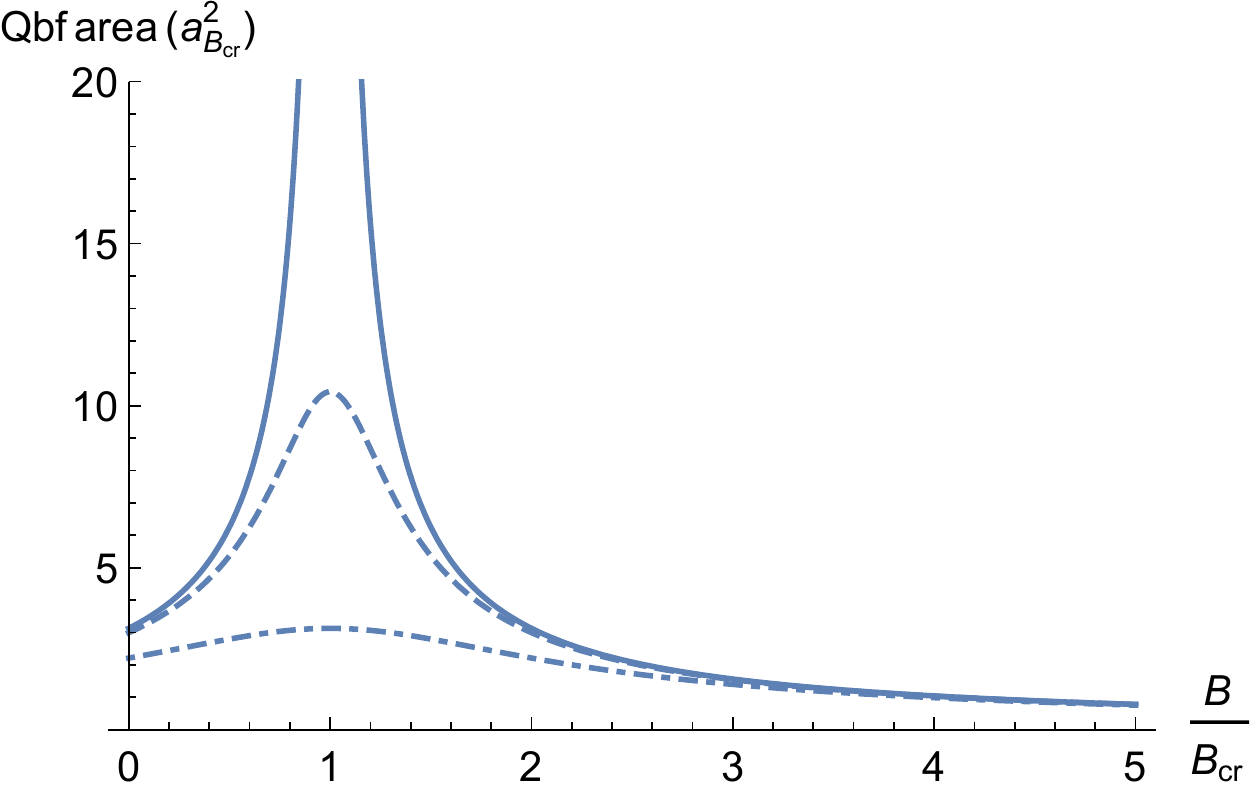}
  \caption{\label{fig:FIG5}Plot of the backflow area 
  {(in units of $a_{B_{cr}}^2$)}, for a state with three components as in Eq.~\protect\eqref{wave1} or as in Eq.~\protect\eqref{psin} with $N=3$, as a function of the magnetic field (in $B_{cr}$ units where $\eta=10^{-25}m_e^2c^2$), for two values of the oscillator frequency ($\omega_1=0.3\, \omega_\eta$ (dashed line), $\omega_2= \omega_\eta$ (dot-dashed line)) compared to the result without oscillator (solid line).}
\end{figure}

 Next we show in Fig.~\ref{fig:FIG5} the area of the quantum backflow regions as function of the external magnetic field expressed in units of the critical magnetic field $B_{\text{cr}}$ given by Eq.~\eqref{bcrit}. In Fig.~\ref{fig:FIG5} the solid line is the value of the area of the quantum backflow region when the oscillator interaction is absent. At $B=B_{\text{cr}}$ the system becomes free and the region where there is backflow is infinite. On the contrary when the oscillator interaction is present the quantity $\beta$ in Eq.~\eqref{abcb} does not vanish and thus the system is always bounded and the backflow region is always finite. The larger the oscillator frequency $\omega$ the stronger this effect (smaller area) as we can see by comparing in Fig.~\ref{fig:FIG5} the dashed line ($\omega= 0.3\, \omega_\eta$) and the dot-dashed line ($\omega= \omega_\eta$).    

\section{Backflow Probability}\label{bfprob}
In the previous sections we have essentially visualized the amount of quantum backflow by showing the area in the ($r,\varphi$) plane where the effective angular momentum becomes positive having considered a quantum state built out of components with only non positive values of $m$. We have also compared various scenarios by computing numerically the backflow area. Nevertheless it would be clearly interesting to have a quantitative estimate of the quantum backflow. One of the ways to obtain this estimate is to introduce the concept of probability associated with angular momentum backflow. 
While in the case of one dimensional problems the issue of the backflow probability has been discussed at length~\cite{yearsley2010,yearsley2012,halliwell,berry10}, it has not been studied within the context of a two-dimensional interacting systems. Perhaps this was due to the difficulties stemming from the two-dimensional nature of the problem. 

A particularly useful approach for the computation of the total backflow probability has been proposed in \cite{berry10}. In this approach a quantum state built with only positive momenta relates the total backflow probability, for a one dimensional problem, to the fraction of the $x$ axis where the local wave number $k(x)$ becomes negative. Here we extend this concept to the case of an interacting two dimensional problem. Our state $\Psi(r,\varphi)$ defined in Eq.~(\ref{wave1}) is a linear combination of eigenfunctions (\ref{unrMa}) with non positive values of the angular momentum ($L_z^{\text{can}}$ component) or non positive values of the magnetic quantum number $m$ ($m\le 0$)  . In our problem the analog of the local wave number $k(x)$ of \cite{berry10} is the local angular momentum ($L_z^{\text{can}}$ component) $\ell(\bm{r})=\hbar m_{\text{eff}}(\bm{r}) $. For a system for which the positions $\bm{r}$ are distributed according to a normalizable probability distribution $|\Psi(\bm{r})|^2$, the probability distribution of the angular momentum is:
\begin{equation}
    P(\ell) =\int\!\!\!\!\!\int \,\,|\Psi(\bm{r})|^2\, \delta \Big( \ell(\bm{r})-\ell\Big)\,  d^2\bm{r}
\label{densityl}\,.
\end{equation}
The above relation can be understood noting that $P(\ell)d\ell$ is the probability of having an angular momentum between $\ell$ and $\ell +d\ell$, and this can be estimated by a\-ve\-ra\-ging over the positions $\bm{r}$ for which $\ell(\bm{r})=\ell$ which are distributed according to $|\Psi(\bm{r})|^2$.

It is convenient, in the following,  to define a probability density with respect to the angular momentum (magnetic) quantum number $m=\ell/\hbar$. This is easily done by extracting the constant $\hbar$ from the $\delta$-function and defining $P(m)=\hbar P(\ell)$:
\begin{equation}
    P(m) =\int\!\!\!\!\!\int \,\,|\Psi(\bm{r})|^2 \,\delta \Big( m_{\text{eff}}(\bm{r})-m\Big)\,  d^2\bm{r}\, .
\label{densityPm}
\end{equation}
So that the total backflow probability (probability of having a positive $m$) is obatained as:
\begin{equation}
     P_{\text{backflow}}=\int_0^\infty P(m) \, dm\, .
     \label{totPbckf1}
\end{equation}
Inserting Eq.~(\ref{densityPm}) into Eq.~(\ref{totPbckf1}) and making use of the fact that the Dirac $\delta$-function is the derivative of the Heaviside $\theta$-function we obtain:
\begin{eqnarray}
P_{\text{backflow}} &=& \int\!\!\!\!\!\int |\Psi(\bm{r})|^2\, d^2\bm{r}\!\!\intop_0^{+\infty}\!\! \left\{ -\frac{\partial}{\partial m} \theta\big( m_{\text{eff}} (\bm{r}) -m\big)\right\} dm\nonumber \\
&=& \int\!\!\!\!\!\int |\Psi(\bm{r})|^2\, d^2\bm{r}\!\!\intop_0^{+\infty}\!\! \Big[ - \theta\big( m_{\text{eff}} (\bm{r}) -m\big)\Big]_{m=0}^{m=+\infty} \nonumber\\
&=& \int\!\!\!\!\!\int |\Psi(\bm{r})|^2\, \theta\big( m_{\text{eff}} (\bm{r})\big) \, d^2\bm{r}
\end{eqnarray}
We may finally write the total angular momentum backflow probability as:
\begin{equation}\label{totPbckf2}
    P_{\text{backflow}}=\int_0^\infty\!\!\!\!\int_0^{2\pi}\, \theta\big(m_{\text{eff}}(r,\varphi)\big)\,\, \left|\Psi(r,\varphi) \right|^2\, r\, dr\,d{\varphi},
\end{equation}
where $\Psi(r,\varphi)$ are the normalized eigenfunction. 

We have computed the total angular momentum backflow probability $P_{\text{backflow}}$ evaluating numerically the integral in Eq.~\eqref{totPbckf2} first for an effective angular momentum defined  in  Eq.~\eqref{meff1} for the state as in Eq.~\eqref{wave1} with a number of components $N=3$, and subsequently for states defined as in Eq.~(\ref{psin}) with increasing number of components up to $N=6$. 
\todo[noline,size=\tiny]{Point n.3 }
We have also examined the variation of the probability of quantum backflow with respect to different choices of the  associated weights $c_m$ with a given number of components $N$. The results of the computations are given in Table~\ref{tab:probability} where we see that $P_{\text{backflow}}$ varies somewhat as the number of components of the state, $N$, is increased and, within a fixed $N$,  as  configurations with different  weights $c_m$ are considered.
In some  cases ($N=3,5,6$) it appears that the configurations with $c_0$ the largest weight have a higher probability relative to the configurations where weights other than $c_0$ are largest but we were unable to generalize it. For instance this  does not happen for $N=4$. This may be traced to the fact that the region where $m_{\text{eff}}>0$ is strongly dependent on the different choices of the weights and also on the different values of $N$ as we have explicitly observed. Also, as $N$ increases, the computation becomes  increasingly challenging from the numerical point of view. We have verified  that, as expected, for any state with  given number of components, $N$, and any configurations of the coefficients $c_m$, the backflow probability is independent of the magnetic field $B$ computing $P_{\text{backflow}}$ for two different values of the magnetic field.
\begin{table}[t!] \caption{\label{tab:probability} Total backflow probability  ($P_{\text{backflow}}$) computed numerically via Eq.~\protect\eqref{totPbckf2} with an  accuracy of one part in $10^3$ for $N=3,4,5,6$ and different choices of the coefficients $c_m$ for any given number of components $N$. The states considered here are all with the lowest value of the radial quantum number ($n=0$) and no oscillator interaction ($\omega=0$).   } 
\begin{ruledtabular} \begin{tabular}{lcc}
$N$ & $(c_{0},c_1, \cdots , c_{N-1})$& $P_{\text{backflow}}$\cr
\hline
3 & $\left(\frac{1}{\sqrt{3}}, \frac{1}{\sqrt{3}}, \frac{1}{\sqrt{3}}\right)$ &  $0.049$\\
- & $\left(\frac{1}{\sqrt{2}}, \frac{1}{\sqrt{4}}, \frac{1}{\sqrt{4}}\right)$ &  $0.103$\\
- & $\left(\frac{1}{\sqrt{4}}, \frac{1}{\sqrt{2}}, \frac{1}{\sqrt{4}}\right)$ &  $0.023$\\
- & $\left(\frac{1}{\sqrt{4}}, \frac{1}{\sqrt{4}}, \frac{1}{\sqrt{2}}\right)$ &  $0.032$\\
4 & $\left(\frac{1}{\sqrt{4}}, \frac{1}{\sqrt{4}}, \frac{1}{\sqrt{4}},\frac{1}{\sqrt{4}}\right)$ &  $0.052$\\
- & $\left(\frac{1}{\sqrt{2}}, \frac{1}{\sqrt{6}}, \frac{1}{\sqrt{6}},\frac{1}{\sqrt{6}}\right)$ &  $0.049$\\
- & $\left(\frac{1}{\sqrt{6}}, \frac{1}{\sqrt{2}}, \frac{1}{\sqrt{6}},\frac{1}{\sqrt{6}}\right)$ &  $0.016$\\
- & $\left(\frac{1}{\sqrt{6}}, \frac{1}{\sqrt{6}}, \frac{1}{\sqrt{2}},\frac{1}{\sqrt{6}}\right)$ &  $0.038$\\
- & $\left(\frac{1}{\sqrt{6}}, \frac{1}{\sqrt{6}}, \frac{1}{\sqrt{6}},\frac{1}{\sqrt{2}}\right)$ &  $0.031$\\
5 & $\left(\frac{1}{\sqrt{5}}, \frac{1}{\sqrt{5}}, \frac{1}{\sqrt{5}},\frac{1}{\sqrt{5}},\frac{1}{\sqrt{5}}\right)$ &  $0.051$\\
- & $\left(\frac{1}{\sqrt{2}}, \frac{1}{\sqrt{8}}, \frac{1}{\sqrt{8}},\frac{1}{\sqrt{8}},\frac{1}{\sqrt{8}}\right)$ &  $0.192$\\
- & $\left(\frac{1}{\sqrt{8}}, \frac{1}{\sqrt{2}}, \frac{1}{\sqrt{8}},\frac{1}{\sqrt{8}},\frac{1}{\sqrt{8}}\right)$ &  $0.011$\\
- & $\left(\frac{1}{\sqrt{8}}, \frac{1}{\sqrt{8}}, \frac{1}{\sqrt{2}},\frac{1}{\sqrt{8}},\frac{1}{\sqrt{8}}\right)$ &  $0.016$\\
- & $\left(\frac{1}{\sqrt{8}}, \frac{1}{\sqrt{8}}, \frac{1}{\sqrt{8}},\frac{1}{\sqrt{2}},\frac{1}{\sqrt{8}}\right)$ &  $0.025$\\
- & $\left(\frac{1}{\sqrt{8}}, \frac{1}{\sqrt{8}}, \frac{1}{\sqrt{8}},\frac{1}{\sqrt{8}},\frac{1}{\sqrt{2}}\right)$ &  $0.029$\\
6 & $\left(\frac{1}{\sqrt{6}}, \frac{1}{\sqrt{6}}, \frac{1}{\sqrt{6}},\frac{1}{\sqrt{6}},\frac{1}{\sqrt{6}},\frac{1}{\sqrt{6}}\right)$ &  $0.047$\\
- & $\left(\frac{1}{\sqrt{2}}, \frac{1}{\sqrt{10}}, \frac{1}{\sqrt{10}},\frac{1}{\sqrt{10}},\frac{1}{\sqrt{10}},\frac{1}{\sqrt{10}}\right)$ &  $0.218$\\
- & $\left(\frac{1}{\sqrt{10}}, \frac{1}{\sqrt{2}}, \frac{1}{\sqrt{10}},\frac{1}{\sqrt{10}},\frac{1}{\sqrt{10}},\frac{1}{\sqrt{10}}\right)$ &  $0.014$\\
- & $\left(\frac{1}{\sqrt{10}}, \frac{1}{\sqrt{10}}, \frac{1}{\sqrt{2}},\frac{1}{\sqrt{10}},\frac{1}{\sqrt{10}},\frac{1}{\sqrt{10}}\right)$&  $0.011$\\
- & $\left(\frac{1}{\sqrt{10}}, \frac{1}{\sqrt{10}}, \frac{1}{\sqrt{10}},\frac{1}{\sqrt{2}},\frac{1}{\sqrt{10}},\frac{1}{\sqrt{10}}\right)$ &  $0.020$\\
- & $\left(\frac{1}{\sqrt{10}}, \frac{1}{\sqrt{10}}, \frac{1}{\sqrt{10}},\frac{1}{\sqrt{10}},\frac{1}{\sqrt{2}},\frac{1}{\sqrt{10}}\right)$ &  $0.023$\\
- & $\left(\frac{1}{\sqrt{10}}, \frac{1}{\sqrt{10}}, \frac{1}{\sqrt{10}},\frac{1}{\sqrt{10}},\frac{1}{\sqrt{10}},\frac{1}{\sqrt{2}}\right)$ &  $0.023$
\end{tabular} 
\end{ruledtabular} 
\end{table}
In retrospect the independence of the total angular momentum backflow probability from the external magnetic field can be understood also from Eq.~\eqref{current} and Eq.~\eqref{meff}. Indeed we see that the term $\bm{j}_2$ in the total current, which depends explicitly on the magnetic field (vector potential) does not contribute to $m_{\text{eff}}$. 
On the other hand from the explicit expression of the exact wave-functions (c.f. Eq.~\eqref{unrM}) the only other dependence on the magnetic field is through the quantity $a_B$. Given that the wave functions depend  only on $(r/a_B,\varphi)$ and that $|\Psi(\bm{r})|^2\propto a_B^{-2} f(r/a_B,\varphi)$ it can be easily shown from Eq.~\eqref{totPbckf2} that the total backflow probability is independent of the magnetic field $B$. 
Let us also remark that while the results in Table~\ref{tab:probability} have been computed for $\omega=0$,  choosing an oscillator interaction    $\omega \ne 0$ simply changes  the value of $a_B$ but not $P_{\text{backflow}}$ since we have shown that the probability is $a_B$ independent.

\section{Discussion and  Conclusions}\label{con}
In this article we have studied the  noncommutative ge\-ne\-ra\-lization of the angular momentum backflow problem considered in \cite{strange}. Extending the exact solution of a free charged particle in a homogeneous magnetic field, we have also considered angular momentum backflow when an oscillator interaction is present. Subsequently we also included momentum noncommutativity only so as to avoid gauge invariance issues associated with noncommuting space coordinates \cite{bertolami}. It has been found that the area where angular momentum backflow takes place changes with the magnetic field which in turn depends on the noncommutativity parameter.

Next, we have attempted to give a quantitative description of the angular momentum backflow. We adopted the approach developed in \cite{berry10} where in a one-dimensional problem the total probability of momentum backflow  is related to the fraction of the $x$-axis where the local wave number $k(x) $ is negative (within a state consisting only of components with positive wave numbers).  We therefore extend the above approach \cite{berry10} to our two-dimensional problem of a charged Dirac particle in a magnetic field. More precisely, in analogy with \cite{strange} we considered a (normalized) physical state $\Psi(\bm{r})$ consisting of a linear combination of eigenfunctions with non positive magnetic angular momentum values ($m\le0$), c.f. Eq.~\eqref{wave1} and relate the probability of backflow to an average, over the probability density distribution $|\Psi(\bm{r})|^2$, of the area of the region of the plane where $m_{\text{eff}}\ge0$, c.f. Eq.~\eqref{totPbckf1} and Eq.~\eqref{totPbckf2}. It has been found that total probability of angular momentum backflow remains the same for different values of the magnetic field. In other words, the backflow probability is independent of the magnetic field. 

It might be noticed that our study of the backflow regions is based on the analysis of the quantity $m_{\text{eff}}$ defined in Eq.~(22a,22b) and as such it is not a gauge invariant quantity because only the total curent $\bm{j}$ of Eq.(19) is gauge invariant while $\bm{j}_1$ and $\bm{j_2}$  are gauge variant. So quantities obtained as averages over the effective local angular momentum $m_{\text{eff}}(r,\varphi)$ such as the expectation value of the canonical angular momentum  
$ \langle \psi|L_z^{\text{can}}|\psi\rangle = \int d^2\bm{r}\, m_{\text{eff}}(r,\varphi) \,|\psi|^2$ would be gauge variant. This might worry the reader that our conclusions  on the angular momentum backflow depend on the gauge choice. However this is not the case as there is a well known subtlety concerning the gauge invariant definition of the orbital angular momentum in the Landau problem~\cite{Greenshields2014:aa,Berche2020:aa,Wakamatsu2018:aa,Kitadono2019:aa}. Indeed it has been shown  that a gauge invariant orbital angular momentum $L_z$ can be defined in terms of the canonical angular momentum $L_z^{\text{can}}=(\bm{r} \times \bm{p})_z$ via $L_z=L_z^{\text{can}} +  \frac{e}{c} r A_\varphi-\frac{e}{2c}\frac{}{}Br^2$~\cite{Wakamatsu2018:aa}. However it turns out that in the symmetric gauge, used throughout this work, the additional contribution $ \frac{e}{c} r A_\varphi-\frac{e}{2c}\frac{}{}Br^2$ vanishes identically and the expectation values of $L_z^{\text{can}}$, computed with $m_{\text{eff}}$ coincide with the expectation values of the gauge invariant $L_z$. Going to another gauge, for instance the Landau gauge, the change in $\langle L_z^{\text{can}}\rangle $ will be compensated by the change in $\langle \left(\frac{e}{c} r A_\varphi-\frac{e}{2c}\frac{}{}Br^2\right)\rangle$ as to obtain the same result of the symmetric gauge. By the same token other observable quantities computed, in the symmetric gauge from $m_{\text{eff}}$ such as, for instance, the back-flow probability, are gauge invariant.

We have also examined the dependence of the quantum backflow probability with respect to the number of components ($N$) in the wave packet as well as the associated weights $c_m$ finding that it
can reach values as high as $P_{\text{backflow}} \approx 0.2$.
More precisely in the present paper we have considered states of the system which are either ($i$) only a simple sum of the $N$ eigenstates with a fixed value of the radial quantum number ($n=0$) or ($ii$) a sum of $N$  components (again with $n=0$) with various choices of different  weights $c_m$. We have not been able to identify a well defined and general pattern of $P_{\text{backflow}}$ when higher values of the number of components $N$ are considered, both in configurations with  equal and different weights $c_m$.   Other possibilities could of course be considered, but they would go beyond the scope of the present work.   Here  we were  mainly interested in  presenting  a sensible definition of the angular momentum backflow probability for the problem of a charged particle in a constant homogeneous magnetic field, with and without the oscillator interaction, along with a complete and  exact solution, of the same problem, also in the presence of non-commutative coordinates.
\todo[noline,size=\tiny]{Point n.4 }
  One interesting feature that emerges from our analysis is that the angular momenutm backflow probability  in our two-dimensional system reaches, in some cases, values ($P_{\text{backflow}} \approx 0.2$) that are larger than the maximum allowed value found by Braken and Malloy $c_{bm} \approx 0.04$ for a one dimensional free system. This may be quite important in view of  upcoming experimental studies of the quantum backflow.

In conclusion in this paper we have been able to confirm the angular momentum backflow of a non relativistic charged particle in a magnetic field even in the presence of non-comutative coordinates. We also succeeded in defining and computing  explicitly the angular momentum total backflow  probability, $P_{\text{backflow}}$, going beyond the results of ref.~\cite{strange} by extending to a two-dimensional physical system the approach developed in \cite{berry10} for momentum backflow in a one-dimensional system. 

It is the authors' opinion that these findings are of interest for further developments in the subject of quantum backflow. 

\begin{acknowledgments}
    One of the authors (PR) thanks Professor M.V. Berry for useful correspondence. He also thanks INFN Sezione di Perugia and the Department of Physics and Geology of the University of Perugia for support and hospitality.
    \end{acknowledgments}
    

\begin{thebibliography}{44}%
\makeatletter
\providecommand \@ifxundefined [1]{%
 \@ifx{#1\undefined}
}%
\providecommand \@ifnum [1]{%
 \ifnum #1\expandafter \@firstoftwo
 \else \expandafter \@secondoftwo
 \fi
}%
\providecommand \@ifx [1]{%
 \ifx #1\expandafter \@firstoftwo
 \else \expandafter \@secondoftwo
 \fi
}%
\providecommand \natexlab [1]{#1}%
\providecommand \enquote  [1]{``#1''}%
\providecommand \bibnamefont  [1]{#1}%
\providecommand \bibfnamefont [1]{#1}%
\providecommand \citenamefont [1]{#1}%
\providecommand \href@noop [0]{\@secondoftwo}%
\providecommand \href [0]{\begingroup \@sanitize@url \@href}%
\providecommand \@href[1]{\@@startlink{#1}\@@href}%
\providecommand \@@href[1]{\endgroup#1\@@endlink}%
\providecommand \@sanitize@url [0]{\catcode `\\12\catcode `\$12\catcode
  `\&12\catcode `\#12\catcode `\^12\catcode `\_12\catcode `\%12\relax}%
\providecommand \@@startlink[1]{}%
\providecommand \@@endlink[0]{}%
\providecommand \url  [0]{\begingroup\@sanitize@url \@url }%
\providecommand \@url [1]{\endgroup\@href {#1}{\urlprefix }}%
\providecommand \urlprefix  [0]{URL }%
\providecommand \Eprint [0]{\href }%
\providecommand \doibase [0]{http://dx.doi.org/}%
\providecommand \selectlanguage [0]{\@gobble}%
\providecommand \bibinfo  [0]{\@secondoftwo}%
\providecommand \bibfield  [0]{\@secondoftwo}%
\providecommand \translation [1]{[#1]}%
\providecommand \BibitemOpen [0]{}%
\providecommand \bibitemStop [0]{}%
\providecommand \bibitemNoStop [0]{.\EOS\space}%
\providecommand \EOS [0]{\spacefactor3000\relax}%
\providecommand \BibitemShut  [1]{\csname bibitem#1\endcsname}%
\let\auto@bib@innerbib\@empty
\bibitem [{\citenamefont {Allcock}(1969{\natexlab{a}})}]{allcock1}%
  \BibitemOpen
  \bibfield  {author} {\bibinfo {author} {\bibfnamefont {G.~R.}\ \bibnamefont
  {Allcock}},\ }\bibfield  {title} {\enquote {\bibinfo {title} {{The time of
  arrival in quantum mechanics I. Formal considerations}},}\ }\href {\doibase
  https://doi.org/10.1016/0003-4916(69)90251-6} {\bibfield  {journal} {\bibinfo
   {journal} {Ann. Phys.}\ }\textbf {\bibinfo {volume} {53}},\ \bibinfo {pages}
  {253} (\bibinfo {year} {1969}{\natexlab{a}})}\BibitemShut {NoStop}%
\bibitem [{\citenamefont {Allcock}(1969{\natexlab{b}})}]{allcock2}%
  \BibitemOpen
  \bibfield  {author} {\bibinfo {author} {\bibfnamefont {G.~R.}\ \bibnamefont
  {Allcock}},\ }\bibfield  {title} {\enquote {\bibinfo {title} {{The time of
  arrival in quantum mechanics II. The individual measurement}},}\ }\href
  {\doibase https://doi.org/10.1016/0003-4916(69)90252-8} {\bibfield  {journal}
  {\bibinfo  {journal} {Ann. Phys.}\ }\textbf {\bibinfo {volume} {53}},\
  \bibinfo {pages} {286} (\bibinfo {year} {1969}{\natexlab{b}})}\BibitemShut
  {NoStop}%
\bibitem [{\citenamefont {{Allcock}}(1969)}]{allcock3}%
  \BibitemOpen
  \bibfield  {author} {\bibinfo {author} {\bibfnamefont {G.~R.}\ \bibnamefont
  {{Allcock}}},\ }\bibfield  {title} {\enquote {\bibinfo {title} {{The time of
  arrival in quantum mechanics III. The measurement ensemble}},}\ }\href
  {\doibase 10.1016/0003-4916(69)90253-X} {\bibfield  {journal} {\bibinfo
  {journal} {Ann. Phys.}\ }\textbf {\bibinfo {volume} {53}},\ \bibinfo {pages}
  {311} (\bibinfo {year} {1969})}\BibitemShut {NoStop}%
\bibitem [{\citenamefont {Bracken}\ and\ \citenamefont {Melloy}(1994)}]{bm94}%
  \BibitemOpen
  \bibfield  {author} {\bibinfo {author} {\bibfnamefont {A.~J.}\ \bibnamefont
  {Bracken}}\ and\ \bibinfo {author} {\bibfnamefont {G.~F.}\ \bibnamefont
  {Melloy}},\ }\bibfield  {title} {\enquote {\bibinfo {title} {Probability
  backflow and a new dimensionless quantum number},}\ }\href
  {http://stacks.iop.org/0305-4470/27/i=6/a=040} {\bibfield  {journal}
  {\bibinfo  {journal} {J. Phys. A: Math. Gen.}\ }\textbf {\bibinfo {volume}
  {27}},\ \bibinfo {pages} {2197} (\bibinfo {year} {1994})}\BibitemShut
  {NoStop}%
\bibitem [{\citenamefont {Yearsley}\ \emph {et~al.}(2012)\citenamefont
  {Yearsley}, \citenamefont {Halliwell}, \citenamefont {Hartshorn},\ and\
  \citenamefont {Whitby}}]{yearsley2012}%
  \BibitemOpen
  \bibfield  {author} {\bibinfo {author} {\bibfnamefont {J.~M.}\ \bibnamefont
  {Yearsley}}, \bibinfo {author} {\bibfnamefont {J.~J.}\ \bibnamefont
  {Halliwell}}, \bibinfo {author} {\bibfnamefont {R.}~\bibnamefont
  {Hartshorn}}, \ and\ \bibinfo {author} {\bibfnamefont {A.}~\bibnamefont
  {Whitby}},\ }\bibfield  {title} {\enquote {\bibinfo {title} {Analytical
  examples, measurement models, and classical limit of quantum backflow},}\
  }\href {\doibase 10.1103/PhysRevA.86.042116} {\bibfield  {journal} {\bibinfo
  {journal} {Phys. Rev. A}\ }\textbf {\bibinfo {volume} {86}},\ \bibinfo
  {pages} {042116} (\bibinfo {year} {2012})}\BibitemShut {NoStop}%
\bibitem [{Note1()}]{Note1}%
  \BibitemOpen
  \bibinfo {note} {The smoothed quasi-projector $\theta _\sigma (\protect
  \mathaccentV {hat}05E{x}) = \DOTSI \intop \ilimits@ _{0}^{\infty } dy \delta
  _\sigma (\protect \mathaccentV {hat}05E{x}-y)$ is defined in terms a smoothed
  (over a length scale $\sigma $) Dirac-$\delta $: $\delta _\sigma (\protect
  \mathaccentV {hat}05E{x}-y) =\protect \frac {1}{\protect \sqrt {2\pi \sigma
  ^2}} \protect \qopname \relax o{exp}\left [-\protect \frac {(\protect
  \mathaccentV {hat}05E{x}-y)^2}{\sigma ^2}\right ]$.}\BibitemShut {Stop}%
\bibitem [{\citenamefont {Melloy}\ and\ \citenamefont
  {Bracken}(1998{\natexlab{a}})}]{melloy2}%
  \BibitemOpen
  \bibfield  {author} {\bibinfo {author} {\bibfnamefont {G.~F.}\ \bibnamefont
  {Melloy}}\ and\ \bibinfo {author} {\bibfnamefont {A.~J.}\ \bibnamefont
  {Bracken}},\ }\bibfield  {title} {\enquote {\bibinfo {title} {{Probability
  Backflow for a Dirac Particle}},}\ }\href@noop {} {\bibfield  {journal}
  {\bibinfo  {journal} {Found. Phys.}\ }\textbf {\bibinfo {volume} {28}},\
  \bibinfo {pages} {505} (\bibinfo {year} {1998}{\natexlab{a}})}\BibitemShut
  {NoStop}%
\bibitem [{\citenamefont {Melloy}\ and\ \citenamefont
  {Bracken}(1998{\natexlab{b}})}]{melloy3}%
  \BibitemOpen
  \bibfield  {author} {\bibinfo {author} {\bibfnamefont {G.F.}\ \bibnamefont
  {Melloy}}\ and\ \bibinfo {author} {\bibfnamefont {A.J.}\ \bibnamefont
  {Bracken}},\ }\bibfield  {title} {\enquote {\bibinfo {title} {The velocity of
  probability transport in quantum mechanics},}\ }\href@noop {} {\bibfield
  {journal} {\bibinfo  {journal} {Annalen der Physik}\ }\textbf {\bibinfo
  {volume} {7}},\ \bibinfo {pages} {726--731} (\bibinfo {year}
  {1998}{\natexlab{b}})},\ \Eprint
  {http://arxiv.org/abs/https://onlinelibrary.wiley.com/doi/pdf/10.1002/}
  {https://onlinelibrary.wiley.com/doi/pdf/10.1002/} \BibitemShut {NoStop}%
\bibitem [{\citenamefont {Su}\ and\ \citenamefont {Chen}(2018)}]{Su:2018aa}%
  \BibitemOpen
  \bibfield  {author} {\bibinfo {author} {\bibfnamefont {Hong-Yi}\ \bibnamefont
  {Su}}\ and\ \bibinfo {author} {\bibfnamefont {Jing-Ling}\ \bibnamefont
  {Chen}},\ }\bibfield  {title} {\enquote {\bibinfo {title} {Quantum backflow
  in solutions to the dirac equation of the spin-1 2 free particle},}\ }\href
  {\doibase 10.1142/S0217732318501869} {\bibfield  {journal} {\bibinfo
  {journal} {Modern Physics Letters A}\ }\textbf {\bibinfo {volume} {33}},\
  \bibinfo {pages} {1850186} (\bibinfo {year} {2018})},\ \Eprint
  {http://arxiv.org/abs/https://doi.org/10.1142/S0217732318501869}
  {https://doi.org/10.1142/S0217732318501869} \BibitemShut {NoStop}%
\bibitem [{\citenamefont {Ashfaque}\ \emph {et~al.}(2019)\citenamefont
  {Ashfaque}, \citenamefont {Lynch},\ and\ \citenamefont
  {Strange}}]{Ashfaque:2019aa}%
  \BibitemOpen
  \bibfield  {author} {\bibinfo {author} {\bibfnamefont {J}~\bibnamefont
  {Ashfaque}}, \bibinfo {author} {\bibfnamefont {J}~\bibnamefont {Lynch}}, \
  and\ \bibinfo {author} {\bibfnamefont {P}~\bibnamefont {Strange}},\
  }\bibfield  {title} {\enquote {\bibinfo {title} {Relativistic quantum
  backflow},}\ }\href {\doibase 10.1088/1402-4896/ab265c} {\bibfield  {journal}
  {\bibinfo  {journal} {Physica Scripta}\ }\textbf {\bibinfo {volume} {94}},\
  \bibinfo {pages} {125107} (\bibinfo {year} {2019})}\BibitemShut {NoStop}%
\bibitem [{\citenamefont {S.~P.~{Eveson}}\ and\ \citenamefont
  {{Verch}}(2005)}]{eveson}%
  \BibitemOpen
  \bibfield  {author} {\bibinfo {author} {\bibfnamefont {C.~J.~{Fewster}}\
  \bibnamefont {S.~P.~{Eveson}}}\ and\ \bibinfo {author} {\bibfnamefont
  {R.}~\bibnamefont {{Verch}}},\ }\bibfield  {title} {\enquote {\bibinfo
  {title} {{Quantum Inequalities in Quantum Mechanics}},}\ }\href {\doibase
  10.1007/s00023-005-0197-9} {\bibfield  {journal} {\bibinfo  {journal}
  {Annales Henri Poincar{\'e}}\ }\textbf {\bibinfo {volume} {6}},\ \bibinfo
  {pages} {1} (\bibinfo {year} {2005})}\BibitemShut {NoStop}%
\bibitem [{\citenamefont {Penz}\ \emph {et~al.}(2006)\citenamefont {Penz},
  \citenamefont {Grübl}, \citenamefont {Kreidl},\ and\ \citenamefont
  {Wagner}}]{penz2005}%
  \BibitemOpen
  \bibfield  {author} {\bibinfo {author} {\bibfnamefont {M.}~\bibnamefont
  {Penz}}, \bibinfo {author} {\bibfnamefont {G.}~\bibnamefont {Grübl}},
  \bibinfo {author} {\bibfnamefont {S.}~\bibnamefont {Kreidl}}, \ and\ \bibinfo
  {author} {\bibfnamefont {P.}~\bibnamefont {Wagner}},\ }\bibfield  {title}
  {\enquote {\bibinfo {title} {A new approach to quantum backflow},}\ }\href
  {http://stacks.iop.org/0305-4470/39/i=2/a=012} {\bibfield  {journal}
  {\bibinfo  {journal} {J. Phys. A: Math. Gen.}\ }\textbf {\bibinfo {volume}
  {39}},\ \bibinfo {pages} {423} (\bibinfo {year} {2006})}\BibitemShut
  {NoStop}%
\bibitem [{\citenamefont {Berry}(2010)}]{berry10}%
  \BibitemOpen
  \bibfield  {author} {\bibinfo {author} {\bibfnamefont {M.~V.}\ \bibnamefont
  {Berry}},\ }\bibfield  {title} {\enquote {\bibinfo {title} {Quantum backflow,
  negative kinetic energy, and optical retro-propagation},}\ }\href
  {http://stacks.iop.org/1751-8121/43/i=41/a=415302} {\bibfield  {journal}
  {\bibinfo  {journal} {J. Phys. A: Math. Theor.}\ }\textbf {\bibinfo {volume}
  {43}},\ \bibinfo {pages} {415302} (\bibinfo {year} {2010})}\BibitemShut
  {NoStop}%
\bibitem [{\citenamefont {van Dijk}\ and\ \citenamefont
  {Toyama}(2019)}]{vanDijk2019aa}%
  \BibitemOpen
  \bibfield  {author} {\bibinfo {author} {\bibfnamefont {W.}~\bibnamefont {van
  Dijk}}\ and\ \bibinfo {author} {\bibfnamefont {F.~M.}\ \bibnamefont
  {Toyama}},\ }\bibfield  {title} {\enquote {\bibinfo {title} {Decay of a
  quasistable quantum system and quantum backflow},}\ }\href {\doibase
  10.1103/PhysRevA.100.052101} {\bibfield  {journal} {\bibinfo  {journal}
  {Phys. Rev. A}\ }\textbf {\bibinfo {volume} {100}},\ \bibinfo {pages}
  {052101} (\bibinfo {year} {2019})}\BibitemShut {NoStop}%
\bibitem [{\citenamefont {Barbier}(2020)}]{Barbier2020:aa}%
  \BibitemOpen
  \bibfield  {author} {\bibinfo {author} {\bibfnamefont {Maximilien}\
  \bibnamefont {Barbier}},\ }\bibfield  {title} {\enquote {\bibinfo {title}
  {Quantum backflow for many-particle systems},}\ }\href {\doibase
  10.1103/PhysRevA.102.023334} {\bibfield  {journal} {\bibinfo  {journal}
  {Phys. Rev. A}\ }\textbf {\bibinfo {volume} {102}},\ \bibinfo {pages}
  {023334} (\bibinfo {year} {2020})}\BibitemShut {NoStop}%
\bibitem [{\citenamefont {Goussev}(2019)}]{Goussev2019:aa}%
  \BibitemOpen
  \bibfield  {author} {\bibinfo {author} {\bibfnamefont {A.}~\bibnamefont
  {Goussev}},\ }\bibfield  {title} {\enquote {\bibinfo {title} {Equivalence
  between quantum backflow and classically forbidden probability flow in a
  diffraction-in-time problem},}\ }\href {\doibase 10.1103/PhysRevA.99.043626}
  {\bibfield  {journal} {\bibinfo  {journal} {Phys. Rev. A}\ }\textbf {\bibinfo
  {volume} {99}},\ \bibinfo {pages} {043626} (\bibinfo {year}
  {2019})}\BibitemShut {NoStop}%
\bibitem [{\citenamefont {Goussev}(2020{\natexlab{a}})}]{goussev:2020aa}%
  \BibitemOpen
  \bibfield  {author} {\bibinfo {author} {\bibfnamefont {Arseni}\ \bibnamefont
  {Goussev}},\ }\bibfield  {title} {\enquote {\bibinfo {title} {Probability
  backflow for correlated quantum states},}\ }\href {\doibase
  10.1103/PhysRevResearch.2.033206} {\bibfield  {journal} {\bibinfo  {journal}
  {Phys. Rev. Research}\ }\textbf {\bibinfo {volume} {2}},\ \bibinfo {pages}
  {033206} (\bibinfo {year} {2020}{\natexlab{a}})}\BibitemShut {NoStop}%
\bibitem [{\citenamefont {Bostelmann}\ \emph {et~al.}(2017)\citenamefont
  {Bostelmann}, \citenamefont {Cadamuro},\ and\ \citenamefont
  {Lechner}}]{Bostelmann:2017aa}%
  \BibitemOpen
  \bibfield  {author} {\bibinfo {author} {\bibfnamefont {Henning}\ \bibnamefont
  {Bostelmann}}, \bibinfo {author} {\bibfnamefont {Daniela}\ \bibnamefont
  {Cadamuro}}, \ and\ \bibinfo {author} {\bibfnamefont {Gandalf}\ \bibnamefont
  {Lechner}},\ }\bibfield  {title} {\enquote {\bibinfo {title} {Quantum
  backflow and scattering},}\ }\href {\doibase 10.1103/PhysRevA.96.012112}
  {\bibfield  {journal} {\bibinfo  {journal} {Phys. Rev. A}\ }\textbf {\bibinfo
  {volume} {96}},\ \bibinfo {pages} {012112} (\bibinfo {year}
  {2017})}\BibitemShut {NoStop}%
\bibitem [{\citenamefont {Strange}(2012)}]{strange}%
  \BibitemOpen
  \bibfield  {author} {\bibinfo {author} {\bibfnamefont {P.}~\bibnamefont
  {Strange}},\ }\bibfield  {title} {\enquote {\bibinfo {title} {Large quantum
  probability backflow and the azimuthal angle–angular momentum uncertainty
  relation for an electron in a constant magnetic field},}\ }\href {\doibase
  10.1088/0143-0807/33/5/1147} {\bibfield  {journal} {\bibinfo  {journal} {Eur.
  J. Phys.}\ }\textbf {\bibinfo {volume} {33}},\ \bibinfo {pages} {1147}
  (\bibinfo {year} {2012})}\BibitemShut {NoStop}%
\bibitem [{\citenamefont {Goussev}(2020{\natexlab{b}})}]{Goussev2020:ab}%
  \BibitemOpen
  \bibfield  {author} {\bibinfo {author} {\bibfnamefont {Arseni}\ \bibnamefont
  {Goussev}},\ }\href@noop {} {\enquote {\bibinfo {title} {Quantum backflow in
  a ring},}\ } (\bibinfo {year} {2020}{\natexlab{b}}),\ \Eprint
  {http://arxiv.org/abs/2008.08022} {arXiv:2008.08022 [quant-ph]} \BibitemShut
  {NoStop}%
\bibitem [{\citenamefont {Douglas}\ and\ \citenamefont
  {Nekrasov}(2001)}]{Douglas:2001aa}%
  \BibitemOpen
  \bibfield  {author} {\bibinfo {author} {\bibfnamefont {M.~R.}\ \bibnamefont
  {Douglas}}\ and\ \bibinfo {author} {\bibfnamefont {N.~A.}\ \bibnamefont
  {Nekrasov}},\ }\bibfield  {title} {\enquote {\bibinfo {title} {Noncommutative
  field theory},}\ }\href {\doibase 10.1103/RevModPhys.73.977} {\bibfield
  {journal} {\bibinfo  {journal} {Rev. Mod. Phys.}\ }\textbf {\bibinfo {volume}
  {73}},\ \bibinfo {pages} {977} (\bibinfo {year} {2001})}\BibitemShut
  {NoStop}%
\bibitem [{\citenamefont {A.~Connes}\ and\ \citenamefont
  {Schwarz}(1998)}]{Connes:1998aa}%
  \BibitemOpen
  \bibfield  {author} {\bibinfo {author} {\bibfnamefont {M.~R.~Douglas}\
  \bibnamefont {A.~Connes}}\ and\ \bibinfo {author} {\bibfnamefont
  {A.}~\bibnamefont {Schwarz}},\ }\bibfield  {title} {\enquote {\bibinfo
  {title} {Noncommutative geometry and matrix theory},}\ }\href
  {http://stacks.iop.org/1126-6708/1998/i=02/a=003} {\bibfield  {journal}
  {\bibinfo  {journal} {J. High Energy Phys.}\ }\textbf {\bibinfo {volume}
  {9802}},\ \bibinfo {pages} {003} (\bibinfo {year} {1998})}\BibitemShut
  {NoStop}%
\bibitem [{\citenamefont {Seiberg}\ and\ \citenamefont
  {Witten}(1999)}]{Seiberg:1999aa}%
  \BibitemOpen
  \bibfield  {author} {\bibinfo {author} {\bibfnamefont {N.}~\bibnamefont
  {Seiberg}}\ and\ \bibinfo {author} {\bibfnamefont {E.}~\bibnamefont
  {Witten}},\ }\bibfield  {title} {\enquote {\bibinfo {title} {String theory
  and noncommutative geometry},}\ }\href
  {http://stacks.iop.org/1126-6708/1999/i=09/a=032} {\bibfield  {journal}
  {\bibinfo  {journal} {J. High Energy Phys.}\ }\textbf {\bibinfo {volume}
  {9909}},\ \bibinfo {pages} {032} (\bibinfo {year} {1999})}\BibitemShut
  {NoStop}%
\bibitem [{\citenamefont {Bellucci}\ \emph {et~al.}(2001)\citenamefont
  {Bellucci}, \citenamefont {Nersessian},\ and\ \citenamefont
  {Sochichiu}}]{Bellucci:2001aa}%
  \BibitemOpen
  \bibfield  {author} {\bibinfo {author} {\bibfnamefont {S.}~\bibnamefont
  {Bellucci}}, \bibinfo {author} {\bibfnamefont {A.}~\bibnamefont
  {Nersessian}}, \ and\ \bibinfo {author} {\bibfnamefont {C.}~\bibnamefont
  {Sochichiu}},\ }\bibfield  {title} {\enquote {\bibinfo {title} {Two phases of
  the noncommutative quantum mechanics},}\ }\href {\doibase
  http://dx.doi.org/10.1016/S0370-2693(01)01304-1} {\bibfield  {journal}
  {\bibinfo  {journal} {Phys. Lett. B}\ }\textbf {\bibinfo {volume} {522}},\
  \bibinfo {pages} {345} (\bibinfo {year} {2001})}\BibitemShut {NoStop}%
\bibitem [{\citenamefont {Smailagic}\ and\ \citenamefont
  {Spallucci}(2002)}]{Smailagic:2002ab}%
  \BibitemOpen
  \bibfield  {author} {\bibinfo {author} {\bibfnamefont {A.}~\bibnamefont
  {Smailagic}}\ and\ \bibinfo {author} {\bibfnamefont {E.}~\bibnamefont
  {Spallucci}},\ }\bibfield  {title} {\enquote {\bibinfo {title}
  {Noncommutative 3d harmonic oscillator},}\ }\href
  {http://stacks.iop.org/0305-4470/35/i=26/a=103} {\bibfield  {journal}
  {\bibinfo  {journal} {J. Phys. A: Math. Gen.}\ }\textbf {\bibinfo {volume}
  {35}},\ \bibinfo {pages} {L363} (\bibinfo {year} {2002})}\BibitemShut
  {NoStop}%
\bibitem [{\citenamefont {J.~Gamboa}\ and\ \citenamefont
  {Rojas}(2002)}]{Gamboa:2002aa}%
  \BibitemOpen
  \bibfield  {author} {\bibinfo {author} {\bibfnamefont {M.~Loewe.}\
  \bibnamefont {J.~Gamboa}, \bibfnamefont {F.~M\'{e}ndez}}\ and\ \bibinfo
  {author} {\bibfnamefont {J.~C.}\ \bibnamefont {Rojas}},\ }\bibfield  {title}
  {\enquote {\bibinfo {title} {Noncommutative quantum mechanics: The
  two-dimensional central field},}\ }\href {\doibase 10.1142/S0217751X02010960}
  {\bibfield  {journal} {\bibinfo  {journal} {Int. J. Mod. Phys. A}\ }\textbf
  {\bibinfo {volume} {17}},\ \bibinfo {pages} {2555} (\bibinfo {year}
  {2002})}\BibitemShut {NoStop}%
\bibitem [{\citenamefont {J.~Gamboa}\ and\ \citenamefont
  {Rojas}(2001)}]{Gamboa:2001ab}%
  \BibitemOpen
  \bibfield  {author} {\bibinfo {author} {\bibfnamefont {M.~Loewe}\
  \bibnamefont {J.~Gamboa}, \bibfnamefont {F.~Mendez}}\ and\ \bibinfo {author}
  {\bibfnamefont {J.~C.}\ \bibnamefont {Rojas}},\ }\bibfield  {title} {\enquote
  {\bibinfo {title} {The landau problem and noncommutative quantum
  mechanics},}\ }\href {\doibase 10.1142/S0217732301005345} {\bibfield
  {journal} {\bibinfo  {journal} {Mod. Phys. Lett. A}\ }\textbf {\bibinfo
  {volume} {16}},\ \bibinfo {pages} {2075} (\bibinfo {year}
  {2001})}\BibitemShut {NoStop}%
\bibitem [{\citenamefont {Chaichian}\ \emph {et~al.}(2001)\citenamefont
  {Chaichian}, \citenamefont {Sheikh-Jabbari},\ and\ \citenamefont
  {Tureanu}}]{PhysRevLett.86.2716}%
  \BibitemOpen
  \bibfield  {author} {\bibinfo {author} {\bibfnamefont {M.}~\bibnamefont
  {Chaichian}}, \bibinfo {author} {\bibfnamefont {M.~M.}\ \bibnamefont
  {Sheikh-Jabbari}}, \ and\ \bibinfo {author} {\bibfnamefont {A.}~\bibnamefont
  {Tureanu}},\ }\bibfield  {title} {\enquote {\bibinfo {title} {Hydrogen atom
  spectrum and the lamb shift in noncommutative {\uppercase{qed}}},}\ }\href
  {\doibase 10.1103/PhysRevLett.86.2716} {\bibfield  {journal} {\bibinfo
  {journal} {Phys. Rev. Lett.}\ }\textbf {\bibinfo {volume} {86}},\ \bibinfo
  {pages} {2716--2719} (\bibinfo {year} {2001})}\BibitemShut {NoStop}%
\bibitem [{\citenamefont {M.~Chaichian}\ and\ \citenamefont
  {Tureanu}(2004)}]{Chaichian2004}%
  \BibitemOpen
  \bibfield  {author} {\bibinfo {author} {\bibfnamefont {M.~M. Sheikh-Jabbari}\
  \bibnamefont {M.~Chaichian}}\ and\ \bibinfo {author} {\bibfnamefont
  {A.}~\bibnamefont {Tureanu}},\ }\bibfield  {title} {\enquote {\bibinfo
  {title} {Non-commutativity of space-time and the hydrogen atom spectrum},}\
  }\href {\doibase 10.1140/epjc/s2004-01886-1} {\bibfield  {journal} {\bibinfo
  {journal} {The European Physical Journal C - Particles and Fields}\ }\textbf
  {\bibinfo {volume} {36}},\ \bibinfo {pages} {251} (\bibinfo {year}
  {2004})}\BibitemShut {NoStop}%
\bibitem [{\citenamefont {Panella}\ and\ \citenamefont
  {Roy}(2014)}]{PhysRevA.90.042111}%
  \BibitemOpen
  \bibfield  {author} {\bibinfo {author} {\bibfnamefont {O.}~\bibnamefont
  {Panella}}\ and\ \bibinfo {author} {\bibfnamefont {P.}~\bibnamefont {Roy}},\
  }\bibfield  {title} {\enquote {\bibinfo {title} {Quantum phase transitions in
  the noncommutative dirac oscillator},}\ }\href {\doibase
  10.1103/PhysRevA.90.042111} {\bibfield  {journal} {\bibinfo  {journal} {Phys.
  Rev. A}\ }\textbf {\bibinfo {volume} {90}},\ \bibinfo {pages} {042111}
  (\bibinfo {year} {2014})}\BibitemShut {NoStop}%
\bibitem [{\citenamefont {Bertolami}\ and\ \citenamefont
  {Queiroz}(2011)}]{bertolami}%
  \BibitemOpen
  \bibfield  {author} {\bibinfo {author} {\bibfnamefont {O.}~\bibnamefont
  {Bertolami}}\ and\ \bibinfo {author} {\bibfnamefont {R.}~\bibnamefont
  {Queiroz}},\ }\bibfield  {title} {\enquote {\bibinfo {title} {Phase-space
  noncommutativity and the {D}irac equation},}\ }\href {\doibase
  https://doi.org/10.1016/j.physleta.2011.09.053} {\bibfield  {journal}
  {\bibinfo  {journal} {Phys. Lett. A}\ }\textbf {\bibinfo {volume} {375}},\
  \bibinfo {pages} {4116} (\bibinfo {year} {2011})}\BibitemShut {NoStop}%
\bibitem [{\citenamefont {Duval}\ and\ \citenamefont
  {Horv{\'{a}}thy}(2001)}]{Duval_2001}%
  \BibitemOpen
  \bibfield  {author} {\bibinfo {author} {\bibfnamefont {C.}~\bibnamefont
  {Duval}}\ and\ \bibinfo {author} {\bibfnamefont {P.~A.}\ \bibnamefont
  {Horv{\'{a}}thy}},\ }\bibfield  {title} {\enquote {\bibinfo {title} {Exotic
  galilean symmetry in the non-commutative plane and the hall effect},}\ }\href
  {\doibase 10.1088/0305-4470/34/47/314} {\bibfield  {journal} {\bibinfo
  {journal} {Journal of Physics A: Mathematical and General}\ }\textbf
  {\bibinfo {volume} {34}},\ \bibinfo {pages} {10097} (\bibinfo {year}
  {2001})}\BibitemShut {NoStop}%
\bibitem [{\citenamefont {Dayi}\ and\ \citenamefont
  {Jellal}(2002)}]{doi:10.1063/1.1504484}%
  \BibitemOpen
  \bibfield  {author} {\bibinfo {author} {\bibfnamefont {O.~F.}\ \bibnamefont
  {Dayi}}\ and\ \bibinfo {author} {\bibfnamefont {A.}~\bibnamefont {Jellal}},\
  }\bibfield  {title} {\enquote {\bibinfo {title} {Hall effect in
  noncommutative coordinates},}\ }\href {\doibase 10.1063/1.1504484} {\bibfield
   {journal} {\bibinfo  {journal} {Journal of Mathematical Physics}\ }\textbf
  {\bibinfo {volume} {43}},\ \bibinfo {pages} {4592} (\bibinfo {year}
  {2002})}\BibitemShut {NoStop}%
\bibitem [{\citenamefont {Palmero}\ \emph {et~al.}(2013)\citenamefont
  {Palmero}, \citenamefont {Torrontegui}, \citenamefont {Muga},\ and\
  \citenamefont {Modugno}}]{PhysRevA.87.053618}%
  \BibitemOpen
  \bibfield  {author} {\bibinfo {author} {\bibfnamefont {M.}~\bibnamefont
  {Palmero}}, \bibinfo {author} {\bibfnamefont {E.}~\bibnamefont
  {Torrontegui}}, \bibinfo {author} {\bibfnamefont {J.~G.}\ \bibnamefont
  {Muga}}, \ and\ \bibinfo {author} {\bibfnamefont {M.}~\bibnamefont
  {Modugno}},\ }\bibfield  {title} {\enquote {\bibinfo {title} {Detecting
  quantum backflow by the density of a bose-einstein condensate},}\ }\href
  {\doibase 10.1103/PhysRevA.87.053618} {\bibfield  {journal} {\bibinfo
  {journal} {Phys. Rev. A}\ }\textbf {\bibinfo {volume} {87}},\ \bibinfo
  {pages} {053618} (\bibinfo {year} {2013})}\BibitemShut {NoStop}%
\bibitem [{\citenamefont {Eliezer}\ \emph {et~al.}(2020)\citenamefont
  {Eliezer}, \citenamefont {Zacharias},\ and\ \citenamefont
  {Bahabad}}]{eliezer2018observation}%
  \BibitemOpen
  \bibfield  {author} {\bibinfo {author} {\bibfnamefont {Yaniv}\ \bibnamefont
  {Eliezer}}, \bibinfo {author} {\bibfnamefont {Thomas}\ \bibnamefont
  {Zacharias}}, \ and\ \bibinfo {author} {\bibfnamefont {Alon}\ \bibnamefont
  {Bahabad}},\ }\bibfield  {title} {\enquote {\bibinfo {title} {Observation of
  optical backflow},}\ }\href {\doibase 10.1364/OPTICA.371494} {\bibfield
  {journal} {\bibinfo  {journal} {Optica}\ }\textbf {\bibinfo {volume} {7}},\
  \bibinfo {pages} {72--76} (\bibinfo {year} {2020})}\BibitemShut {NoStop}%
\bibitem [{\citenamefont {Bertolami}\ \emph {et~al.}(2005)\citenamefont
  {Bertolami}, \citenamefont {Rosa}, \citenamefont {de~Arag\~ao}, \citenamefont
  {Castorina},\ and\ \citenamefont {Zappal\`a}}]{Bertolami:2005aa}%
  \BibitemOpen
  \bibfield  {author} {\bibinfo {author} {\bibfnamefont {O.}~\bibnamefont
  {Bertolami}}, \bibinfo {author} {\bibfnamefont {J.~G.}\ \bibnamefont {Rosa}},
  \bibinfo {author} {\bibfnamefont {C.~M.~L.}\ \bibnamefont {de~Arag\~ao}},
  \bibinfo {author} {\bibfnamefont {P.}~\bibnamefont {Castorina}}, \ and\
  \bibinfo {author} {\bibfnamefont {D.}~\bibnamefont {Zappal\`a}},\ }\bibfield
  {title} {\enquote {\bibinfo {title} {Noncommutative gravitational quantum
  well},}\ }\href {\doibase 10.1103/PhysRevD.72.025010} {\bibfield  {journal}
  {\bibinfo  {journal} {Phys. Rev. D}\ }\textbf {\bibinfo {volume} {72}},\
  \bibinfo {pages} {025010} (\bibinfo {year} {2005})}\BibitemShut {NoStop}%
\bibitem [{\citenamefont {Kokado}\ \emph {et~al.}(2004)\citenamefont {Kokado},
  \citenamefont {Okamura},\ and\ \citenamefont {Saito}}]{Kokado:2004aa}%
  \BibitemOpen
  \bibfield  {author} {\bibinfo {author} {\bibfnamefont {Akira}\ \bibnamefont
  {Kokado}}, \bibinfo {author} {\bibfnamefont {Takashi}\ \bibnamefont
  {Okamura}}, \ and\ \bibinfo {author} {\bibfnamefont {Takesi}\ \bibnamefont
  {Saito}},\ }\bibfield  {title} {\enquote {\bibinfo {title} {Noncommutative
  quantum mechanics and the {Seiberg-Witten} map},}\ }\href {\doibase
  10.1103/PhysRevD.69.125007} {\bibfield  {journal} {\bibinfo  {journal} {Phys.
  Rev. D}\ }\textbf {\bibinfo {volume} {69}},\ \bibinfo {pages} {125007}
  (\bibinfo {year} {2004})}\BibitemShut {NoStop}%
\bibitem [{\citenamefont {Fl{\"u}gge}(1974)}]{Flugge:1974aa}%
  \BibitemOpen
  \bibfield  {author} {\bibinfo {author} {\bibfnamefont {S.}~\bibnamefont
  {Fl{\"u}gge}},\ }\href@noop {} {\emph {\bibinfo {title} {Practical quantum
  mechanics. {I}, {II}}}}\ (\bibinfo  {publisher} {Springer-Verlag},\ \bibinfo
  {address} {Berlin},\ \bibinfo {year} {1974})\BibitemShut {NoStop}%
\bibitem [{\citenamefont {Yearsley}(2010)}]{yearsley2010}%
  \BibitemOpen
  \bibfield  {author} {\bibinfo {author} {\bibfnamefont {J.~M.}\ \bibnamefont
  {Yearsley}},\ }\bibfield  {title} {\enquote {\bibinfo {title} {Quantum
  arrival time for open systems},}\ }\href {\doibase
  10.1103/PhysRevA.82.012116} {\bibfield  {journal} {\bibinfo  {journal} {Phys.
  Rev. A}\ }\textbf {\bibinfo {volume} {82}},\ \bibinfo {pages} {012116}
  (\bibinfo {year} {2010})}\BibitemShut {NoStop}%
\bibitem [{\citenamefont {Halliwell}\ \emph {et~al.}(2013)\citenamefont
  {Halliwell}, \citenamefont {Gillman}, \citenamefont {Lennon}, \citenamefont
  {Patel},\ and\ \citenamefont {Ramirez}}]{halliwell}%
  \BibitemOpen
  \bibfield  {author} {\bibinfo {author} {\bibfnamefont {J.~J.}\ \bibnamefont
  {Halliwell}}, \bibinfo {author} {\bibfnamefont {E.}~\bibnamefont {Gillman}},
  \bibinfo {author} {\bibfnamefont {O.}~\bibnamefont {Lennon}}, \bibinfo
  {author} {\bibfnamefont {M.}~\bibnamefont {Patel}}, \ and\ \bibinfo {author}
  {\bibfnamefont {I.}~\bibnamefont {Ramirez}},\ }\bibfield  {title} {\enquote
  {\bibinfo {title} {Quantum backflow states from eigenstates of the
  regularized current operator},}\ }\href
  {http://stacks.iop.org/1751-8121/46/i=47/a=475303} {\bibfield  {journal}
  {\bibinfo  {journal} {J. Phys. A: Math. Theor.}\ }\textbf {\bibinfo {volume}
  {46}},\ \bibinfo {pages} {475303} (\bibinfo {year} {2013})}\BibitemShut
  {NoStop}%
\bibitem [{\citenamefont {Greenshields}\ \emph {et~al.}(2014)\citenamefont
  {Greenshields}, \citenamefont {Stamps}, \citenamefont {Franke-Arnold},\ and\
  \citenamefont {Barnett}}]{Greenshields2014:aa}%
  \BibitemOpen
  \bibfield  {author} {\bibinfo {author} {\bibfnamefont {Colin~R.}\
  \bibnamefont {Greenshields}}, \bibinfo {author} {\bibfnamefont {Robert~L.}\
  \bibnamefont {Stamps}}, \bibinfo {author} {\bibfnamefont {Sonja}\
  \bibnamefont {Franke-Arnold}}, \ and\ \bibinfo {author} {\bibfnamefont
  {Stephen~M.}\ \bibnamefont {Barnett}},\ }\bibfield  {title} {\enquote
  {\bibinfo {title} {Is the angular momentum of an electron conserved in a
  uniform magnetic field?}}\ }\href {\doibase 10.1103/PhysRevLett.113.240404}
  {\bibfield  {journal} {\bibinfo  {journal} {Phys. Rev. Lett.}\ }\textbf
  {\bibinfo {volume} {113}},\ \bibinfo {pages} {240404} (\bibinfo {year}
  {2014})}\BibitemShut {NoStop}%
\bibitem [{\citenamefont {Berche}\ \emph {et~al.}(2016)\citenamefont {Berche},
  \citenamefont {Malterre},\ and\ \citenamefont {Medina}}]{Berche2020:aa}%
  \BibitemOpen
  \bibfield  {author} {\bibinfo {author} {\bibfnamefont {Bertrand}\
  \bibnamefont {Berche}}, \bibinfo {author} {\bibfnamefont {Daniel}\
  \bibnamefont {Malterre}}, \ and\ \bibinfo {author} {\bibfnamefont {Ernesto}\
  \bibnamefont {Medina}},\ }\bibfield  {title} {\enquote {\bibinfo {title}
  {Gauge transformations and conserved quantities in classical and quantum
  mechanics},}\ }\href {\doibase 10.1119/1.4955153} {\bibfield  {journal}
  {\bibinfo  {journal} {American Journal of Physics}\ }\textbf {\bibinfo
  {volume} {84}},\ \bibinfo {pages} {616--625} (\bibinfo {year} {2016})},\
  \Eprint {http://arxiv.org/abs/https://doi.org/10.1119/1.4955153}
  {https://doi.org/10.1119/1.4955153} \BibitemShut {NoStop}%
\bibitem [{\citenamefont {Wakamatsu}\ \emph {et~al.}(2018)\citenamefont
  {Wakamatsu}, \citenamefont {Kitadono},\ and\ \citenamefont
  {Zhang}}]{Wakamatsu2018:aa}%
  \BibitemOpen
  \bibfield  {author} {\bibinfo {author} {\bibfnamefont {M.}~\bibnamefont
  {Wakamatsu}}, \bibinfo {author} {\bibfnamefont {Y.}~\bibnamefont {Kitadono}},
  \ and\ \bibinfo {author} {\bibfnamefont {P.-M.}\ \bibnamefont {Zhang}},\
  }\bibfield  {title} {\enquote {\bibinfo {title} {The issue of gauge choice in
  the {L}andau problem and the physics of canonical and mechanical orbital
  angular momenta},}\ }\href {\doibase
  https://doi.org/10.1016/j.aop.2018.03.019} {\bibfield  {journal} {\bibinfo
  {journal} {Annals of Physics}\ }\textbf {\bibinfo {volume} {392}},\ \bibinfo
  {pages} {287 -- 322} (\bibinfo {year} {2018})}\BibitemShut {NoStop}%
\bibitem [{\citenamefont {Kitadono}()}]{Kitadono2019:aa}%
  \BibitemOpen
  \bibfield  {author} {\bibinfo {author} {\bibfnamefont {Yoshio}\ \bibnamefont
  {Kitadono}},\ }\enquote {\bibinfo {title} {{A {T}est of Gauge Invariant
  Canonical Angular Momentum in Landau Level Problem}},}\ in\ \href {\doibase
  10.7566/JPSCP.26.021015} {\emph {\bibinfo {booktitle} {Proceedings of the 8th
  International Conference on Quarks\- and Nuclear Physics
  (QNP2018)}}}\BibitemShut {NoStop}%
\end{thebibliography}
%

\end{document}